\documentclass[final,times,5p]{elsarticle}
\usepackage{ctable}
\usepackage{ftnright}
\usepackage{graphicx}
\usepackage{xspace}
\usepackage{hhline}
\usepackage{amsmath}
\usepackage{amsfonts}
\usepackage{hyperref}

\DeclareMathOperator{\Imag}{Im}
\newcommand{\ee}{\ensuremath{e^{+}e^{-}}\xspace}
\newcommand{\mumu}{\ensuremath{\mu^{+}\mu^{-}}}

\newcommand{\PP}{\ensuremath{\psi(2S)}\xspace}
\newcommand{\JP}{\ensuremath{J/\psi}\xspace}

\renewcommand{\epsilon}{\varepsilon}

\begin{document}
\title{Precise measurement of $R_{\text{uds}}$ and $R$ between 1.84 and 3.72 GeV at the KEDR detector}

\author[binp]{V.V.~Anashin}
\author[binp]{O.V.~Anchugov}
\author[binp,nsu]{V.M.~Aulchenko}
\author[binp,nsu]{E.M.~Baldin}
\author[binp,nstu]{G.N.~Baranov} 
\author[binp]{A.K.~Barladyan}
\author[binp,nsu]{A.Yu.~Barnyakov}
\author[binp,nsu]{M.Yu.~Barnyakov}
\author[binp,nsu]{S.E.~Baru}
\author[binp]{I.Yu.~Basok}
\author[binp]{A.M.~Batrakov}
\author[binp]{E.A.~Bekhtenev}
\author[binp,nsu]{A.E.~Blinov}
\author[binp,nsu,nstu]{V.E.~Blinov}
\author[binp,nsu]{A.V.~Bobrov}
\author[binp,nsu]{V.S.~Bobrovnikov}
\author[binp,nsu]{A.V.~Bogomyagkov}
\author[binp,nsu]{A.E.~Bondar}
\author[binp,nsu]{A.R.~Buzykaev}
\author[binp,nsu]{P.B.~Cheblakov}
\author[binp,nstu]{V.L.~Dorohov}
\author[binp,nsu]{S.I.~Eidelman}
\author[binp,nsu,nstu]{D.N.~Grigoriev}
\author[binp]{S.A.~Glukhov}
\author[binp]{V.V.~Kaminskiy}
\author[binp]{S.E.~Karnaev}
\author[binp]{G.V.~Karpov}
\author[binp]{S.V.~Karpov}
\author[binp,nstu]{K.Yu.~Karukina}
\author[binp]{D.P.~Kashtankin}
\author[binp]{P.V.~Kasyanenko}
\author[binp]{T.A.~Kharlamova}
\author[binp]{V.A.~Kiselev}
\author[binp]{V.V.~Kolmogorov}
\author[binp,nsu]{S.A.~Kononov}
\author[binp]{K.Yu.~Kotov}
\author[binp]{A.A.~Krasnov}
\author[binp,nsu]{E.A.~Kravchenko}
\author[binp,nsu]{V.N.~Kudryavtsev}
\author[binp,nsu]{V.F.~Kulikov}
\author[binp,nstu]{G.Ya.~Kurkin}
\author[binp]{I.A.~Kuyanov}
\author[binp,nstu]{E.B.~Levichev}
\author[binp,nsu]{D.A.~Maksimov}
\author[binp]{V.M.~Malyshev}
\author[binp,nsu]{A.L.~Maslennikov}
\author[binp,nsu]{O.I.~Meshkov}
\author[binp]{S.I.~Mishnev}
\author[binp]{I.A.~Morozov}
\author[binp,nsu]{I.I.~Morozov}
\author[binp]{S.A.~Nikitin}
\author[binp,nsu]{I.B.~Nikolaev}
\author[binp]{I.N.~Okunev}
\author[binp,nsu,nstu]{A.P.~Onuchin}
\author[binp]{S.B.~Oreshkin}
\author[binp,nsu]{A.A.~Osipov}
\author[binp,nstu]{I.V.~Ovtin}
\author[binp,nsu]{S.V.~Peleganchuk}
\author[binp,nstu]{S.G.~Pivovarov}
\author[binp]{P.A.~Piminov}
\author[binp]{V.V.~Petrov}
\author[binp,nsu]{V.G.~Prisekin}
\author[binp,nsu]{O.L.~Rezanova}
\author[binp,nsu]{A.A.~Ruban}
\author[binp]{G.A.~Savinov}
\author[binp,nsu]{A.G.~Shamov}
\author[binp]{D.N.~Shatilov}
\author[binp]{D.A.~Shvedov}
\author[binp,nsu]{B.A.~Shwartz}
\author[binp]{E.A.~Simonov}
\author[binp]{S.V.~Sinyatkin}
\author[binp]{A.N.~Skrinsky}
\author[binp,nsu]{A.V.~Sokolov}
\author[binp]{D.P.~Sukhanov}
\author[binp,nsu]{A.M.~Sukharev}
\author[binp,nsu]{E.V.~Starostina}
\author[binp,nsu]{A.A.~Talyshev}
\author[binp,nsu]{V.A.~Tayursky}
\author[binp,nsu]{V.I.~Telnov}
\author[binp,nsu]{Yu.A.~Tikhonov}
\author[binp,nsu]{K.Yu.~Todyshev \corref{cor}}
\cortext[cor]{Corresponding author, e-mail: todyshev@inp.nsk.su}
\author[binp]{A.G.~Tribendis}
\author[binp]{G.M.~Tumaikin}
\author[binp]{Yu.V.~Usov}
\author[binp]{A.I.~Vorobiov}
\author[binp,nsu]{V.N.~Zhilich}
\author[binp]{A.A.~Zhukov}
\author[binp,nsu]{V.V.~Zhulanov}
\author[binp,nsu]{A.N.~Zhuravlev}
 \address[binp]{Budker Institute of Nuclear Physics, 11, akademika
 Lavrentieva prospect,  Novosibirsk, 630090, Russia}
 \address[nsu]{Novosibirsk State University, 2, Pirogova street,  Novosibirsk, 630090, Russia}
 \address[nstu]{Novosibirsk State Technical University, 20, Karl Marx
  prospect,  Novosibirsk, 630092, Russia}

\begin{abstract} 
The present work continues a series of the KEDR measurements of the $R$
value that started in 2010 at the VEPP-4M \ee collider. 
By combining new data with our previous results in this energy range 
we measured the values of $R_{\text{uds}}$ and $R$ at nine
center-of-mass energies between 3.08 and 3.72 GeV. 
The total accuracy is about or better than  $2.6\%$ at most
of energy points  with a systematic uncertainty of about $1.9\%$.
Together with the previous precise $R$ measurement at KEDR in the energy range
1.84-3.05 GeV, it constitutes the most detailed high-precision $R$ measurement
near the charmonium production threshold.
\end{abstract}
\maketitle
\section{Introduction}\label{sec:intro}
The ratio of the radiatively corrected total cross section of
electron-positron annihilation into hadrons to the lowest-order QED cross
section  of the muon pair production is referred to as the value of $R$.
This quantity plays critical role in various precision tests of the 
Standard Model, e.g. $R(s)$ measurements are employed to determine the
hadronic contribution to the 
anomalous magnetic moment of the muon  and the value of the
electromagnetic fine structure constant at the $Z^0$ peak
$\alpha(M_Z^2)$~\cite{dhmz,hlmnt}, the running strong  coupling 
constant $\alpha_s(s)$ and heavy quark masses~\cite{quark}.
 
More than ten experiments contributed to  the  $R(s)$ measurement  
in the energy range between the $p\bar{p}$ and $D\bar{D}$ thresholds
\cite{ADONEMUPI:R1973,Mark1:R1977,PLUTO:R1977,GG2:R1979,MARK2:R1980,ADONE:R1981,
MARK1:R1982,Bai:1999pk,BES:R2002,BES:R2006,BES:R2009,KEDR:R2016,KEDR:R2017}. 
The most accurate results  were obtained  in the experiments of
BES-II~\cite{BES:R2009} and KEDR~\cite{KEDR:R2016,KEDR:R2017}, 
in which the accuracy of about 3.3\%  was reached at separate points. 

For the considered energy range, systematic uncertainties 
give a substantial contribution to the total accuracy of the  $R(s)$ quantity.
This fact motivated us to repeat the $R$ measurement 
in the given energy range after repairing and upgrading the
detector. In 2014 the region of the $J/\psi$ and $\psi(2S)$ resonances
was scanned in the KEDR experiment with an integrated
luminosity of about 1.3~pb$^{-1}$. 
\section{Experiment}\label{sec:exp}
The experiment was carried out at the VEPP-4M \cite{Anashin:1998sj} 
collider in the same approach that was used earlier in \cite{KEDR:R2016}.

The  KEDR detector and its performance are described in detail 
elsewhere~\cite{KEDR:Det}. 
At the end of 2013, the repair and upgrade of the detector were completed.
The vacuum chamber was replaced with a new wider one 
to reduce possible accelerator background. 
The preamplifiers  of the VD were reconfigured and
equipped with additional copper-foil screens to suppress the crosstalk.
The drift chamber was renovated  and a few layers were repaired.
A second layer of the aerogel Cherenkov counters was installed.
The barrel part of the TOF system was equipped with additional
magnetic shields to suppress the reduction of
signal amplitudes in photomultipliers in the magnetic
field. The entire krypton  was cleaned of electronegative impurities.

The purpose of the experiment was to repeat the $R$ scan  carried out
by KEDR in 2011, in addition we collected data at the
energy point below the $J/\psi$. The total hadronic cross
section was measured at eight  points between 3.08 and 3.72 GeV. 
The value of energy was calculated by interpolating  
the resonance depolarization results obtained in calibration runs. 

\begin{figure*}[t!]
\begin{center}
\includegraphics[width=0.98\textwidth]{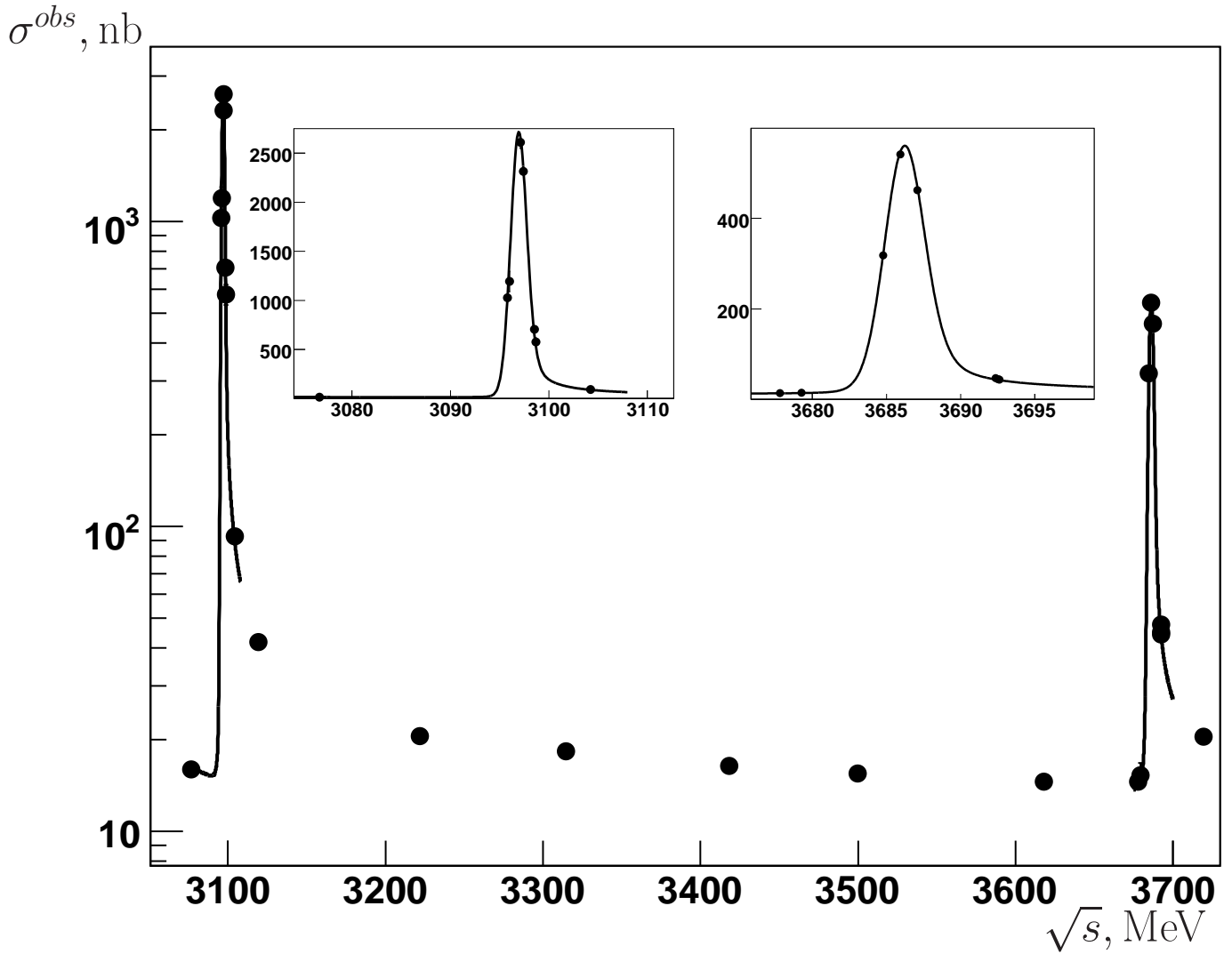}
\caption{{The observed multihadronic cross section as a function 
of the c.m. energy for the two scans. 
The curves are the result of the fits of the narrow resonances.
The insets show closeup of the $J/\psi$ and $\psi(2S)$ regions.
\label{scans_picture}
}}
\end{center}
\end{figure*}

The actual energies  and the integrated luminosity at the points are 
presented in Table~\ref{tab:epoints}. To determine resonance parameters 
additional data samples of about 
0.34~pb$^{-1}$ were taken in the vicinity of the $J/\psi$ and $\psi(2S)$ 
resonances. A measurement of beam energy by the resonance
depolarization method was carried out at least once at each 
listed point off the resonance peak regions.
The assigned energy errors are due to the drift of the parameters
of the accelerator during data taking.
The data points and the resonance fits are shown in  Fig.~\ref{scans_picture}.

\renewcommand{\arraystretch}{1.7}
\setlength{\tabcolsep}{3pt}
\begin{table}[h!]
\begin{center}
\caption{{\label{tab:epoints} Center-of-mass energy $\sqrt{s}$ and  
integrated luminosity $\int\!\!\mathcal{L}dt$.}} 
\begin{tabular}{lccc}
Point & $\sqrt{s}$, MeV &  $\int\!\!\mathcal{L}dt$, nb$^{-1}$  \\  \hline
1&$3076.7 \pm 0.2 $   &       $103.45  \pm 0.98 \pm 0.93$ \\\hline 
2&$3119.2 \pm 0.2 $   &       $~77.15  \pm 0.86 \pm 0.69$ \\\hline 
3&$3221.8 \pm 0.2 $   &       $~93.18  \pm 0.98 \pm 0.84$ \\\hline 
4&$3314.7 \pm 0.4 $   &       $157.69  \pm 1.31 \pm 1.42 $\\\hline 
5&$3418.3 \pm 0.8 $   &       $150.46  \pm 1.33 \pm 1.35 $\\\hline 
6&$3499.6 \pm 1.1 $   &       $125.76  \pm 1.23 \pm 1.13$\\\hline 
7&$3618.1 \pm 0.4 $   &       $159.97  \pm 1.43 \pm 1.44$\\\hline 
8&$3719.6 \pm 0.2 $   &       $130.90  \pm 1.34 \pm 1.18$ \\ \hline
\end{tabular} 
\end{center}
\end{table}
\renewcommand{\arraystretch}{1.}
\section{Data analysis}\label{sec:data}
\subsection{Analysis procedure}\label{subsec:proc}
Details of the analysis procedure are provided in \cite{KEDR:R2016}.




To determine the $R$ value we take into account narrow resonances explicitly 
instead of including them in the radiative correction $\delta(s)$. 
The narrow-resonance cross section  depends on the combination
$\epsilon_{\psi}\Gamma_{ee}\mathcal{B}_{h}$. The efficiencies $\epsilon_{\psi}$ 
were extracted by fitting the data at the resonance regions, thus the 
obtained resonance cross section 
is not sensitive to the world-average values of the leptonic width 
$\Gamma_{ee}$ and the hadronic branching fraction $\mathcal{B}_{h}$ used.
Computations of a narrow-resonance cross section, 
the resonance -- continuum interference and the resonance fitting 
procedure are described in more detail in 
Refs.~\cite{psi2S:2012,MASS::KEDR2015}.

The floating parameters were the detection efficiency $\epsilon_{\psi}$
at the world-average values of the leptonic width $\Gamma_{ee}$ and 
its product by the hadronic
branching fraction $\mathcal{B}_{h}$, the machine energy spread 
and the magnitude of the continuum  cross section observed at the reference 
point below the resonance.  
The $J/\psi$ and $\psi(2S)$ detection efficiencies,
the collision energy spreads obtained and the $\chi^2$ probabilities of 
the fits are presented in Table~\ref{tab:fits}. 
\renewcommand{\arraystretch}{1.2}
\setlength{\tabcolsep}{3pt}
\begin{table}[h!]
\begin{center}
\caption{ \label{tab:fits} { Efficiency, energy spread and $\chi^2$ 
probability  of the fits of the $J/\psi$ and $\psi(2S)$ resonances
(statistical errors only are presented). 
The reference energy points for the energy spread parameters 
correspond to masses of the $J/\psi$ and $\psi(2S)$ mesons taken 
from PDG~\cite{PDG:2014}.}}
\begin{tabular}{cccccc} 
           & Efficiency, $\%$ & $\sigma_W$, MeV &$P(\chi^2)$, $\%$  \\\hline
$J/\psi$   &  $78.72\pm  0.89$ &  $0.785\pm  0.004$  & $53.5$   \\\hline
$\psi(2S)$ &  $80.65\pm  1.95$ &  $1.262\pm  0.045$  & $99.4$  \\ \hline
\end{tabular}
\end{center}
\end{table}

Table \ref{tab:nrc} lists the relative contribution of the
$J/\psi$ and $\psi(2S)$ to the observed cross section. 
\renewcommand{\arraystretch}{1.4}
\begin{table}[h]
\begin{center}
\caption{{\label{tab:nrc} Relative contribution of the $J/\psi$ and
    $\psi(2S)$ resonances to the observed multihadronic cross
    section. Negative signs correspond to resonance -- continuum 
interference.}} 
\begin{tabular}{lll}
Point &  $\frac{\sigma_{J/\psi}}{\sigma_{\text{obs}}}$,$\%$ &$\frac{\sigma_{\psi(2S)}}{\sigma_{\text{obs}}}$,$\%$   \\  \hline
1& $-7.24 (\text{interference})$&\\ \hline 
2&   59.71  & \\ \hline
3&   22.63  &\\ \hline
4&   14.83  & \\ \hline
5&   10.75  & \\ \hline
6&  ~8.76  & \\ \hline
7&  ~6.80  & $-0.76 ({\text{interference}}) $\\\hline 
8&  ~4.05  & 28.27 \\ \hline
\end{tabular} 
\end{center}
\end{table}
\renewcommand{\arraystretch}{1.2}

The detection efficiencies for the single-photon annihilation to hadrons 
$\epsilon(s)$ and background processes were obtained from simulation.

The radiative correction factor is determined by excluding a contribution
of the $J/\psi$ and $\psi(2S)$ resonances and can be written as
\begin{equation}
\label{eq:RadDelta}
1+\delta(s)=\int\!\frac{dx}{1\!-\!x}\, 
\frac{\mathcal{F}(s,x)}{\big|1-\tilde{\Pi}((1\!-\!x)s)\big|^2}\,
\frac{\tilde{R}((1\!-\!x)s)\,\epsilon((1\!-\!x)s)}{R(s)\,\epsilon(s)},
\end{equation}
where $\mathcal{F}(s,x)$ is  the radiative correction kernel~\cite{KF:1985}.
The variable $x$ is a fraction of $s$ lost due to the initial-state
radiation.
The vacuum polarization operator $\tilde{\Pi}$ and  the quantity
$\tilde{R}$  do not include a contribution of narrow resonances, 
details of the calculation are presented in Sec.~\ref{sec:radeff}.

Thus, we extract the $R_{\text{uds}}$ value, then by adding the contribution 
of narrow resonances we obtain the quantity $R$.

\subsection{Monte Carlo simulation}\label{sec:MC}
The KEDR simulation program is based on the GEANT package, 
version 3.21~\cite{GEANT:Tool}.

To simulate single-photon annihilation to hadrons we employ the JETSET~7.4 
code~\cite{JETSET} with the parameters tuned 
at each energy point. As an alternative way of generating events of 
the $\text{uds}$ continuum,  we use the LUARLW generator~\cite{LUARLW:2001}.

Bhabha scattering events required for the precise luminosity determination are
simulated by BHWIDE~\cite{BHWIDEGEN}. The MCGPJ generator~\cite{MCGPJ}
provides simulation of $\mu^{+}\mu^{-}$ events and the  $\ee\to\ee\gamma$
process as an alternative to BHWIDE. 
The detection efficiency for   $\tau^{+}\tau^{-}$ 
events was obtained using the KORALB event generator~\cite{KORALB24}.
The two-photon processes $e^{+}e^{-}\to e^{+}e^{-} X$ are simulated with 
the generators described in Refs.~\cite{BERENDS:EEEE,BERENDS:EEMM,KEDR:EEX}.

The \JP and \PP decays were simulated with the tuned version of the 
BES generator~\cite{BESGEN} based on the JETSET~7.4 code
~\cite{psi2S:2012,jpsi:2018}.

During the whole experiment random trigger events were recorded.
These events were embedded into the Monte Carlo simulated data 
to account for various detector noises and a coincidence of the 
simulated processes with the collider and cosmic backgrounds.

\begin{center}
\begin{figure*}[!ht]
\begin{center}
\includegraphics[width=0.4\textwidth,height=0.25\textheight]{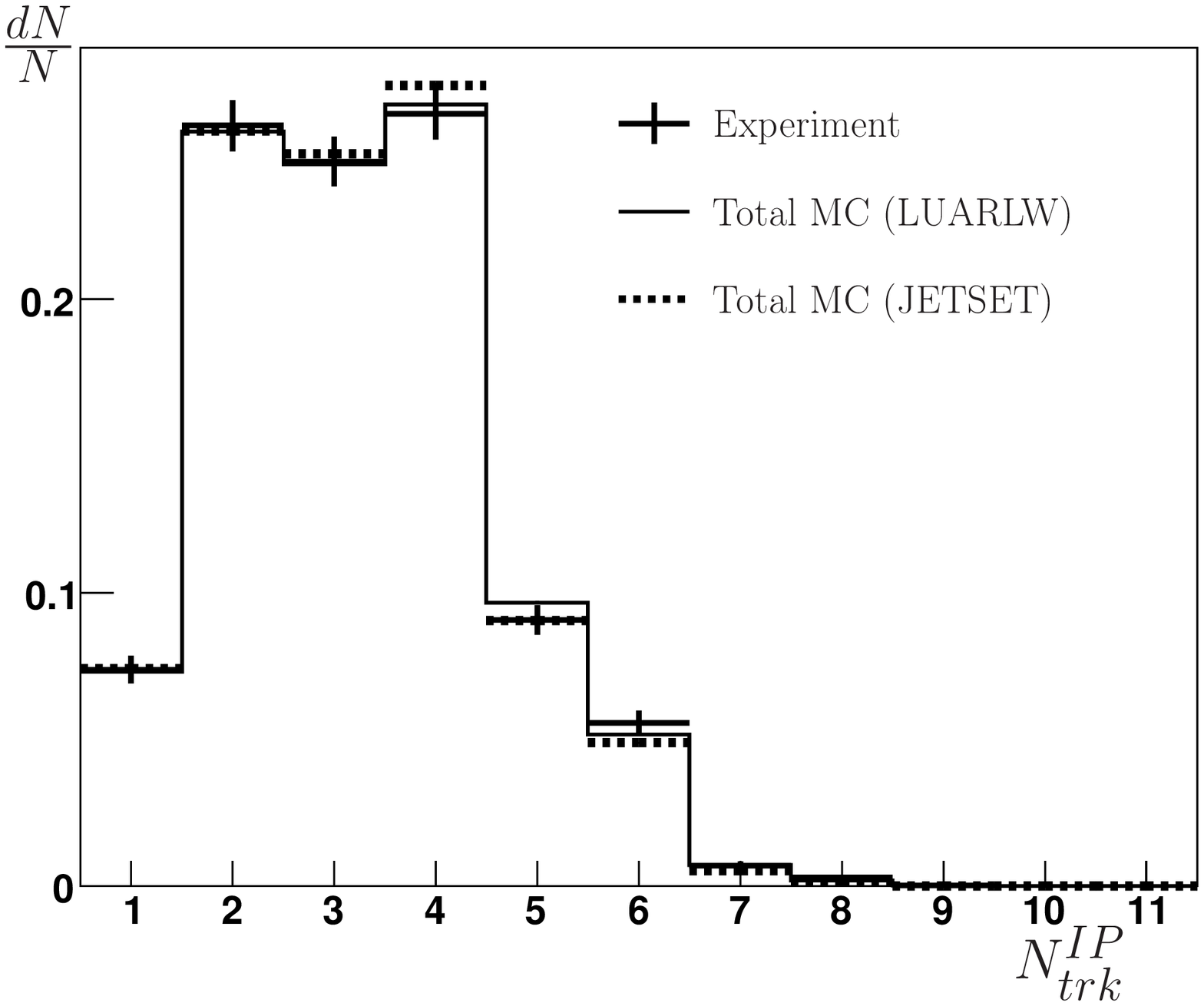}
\hspace*{0.2cm}
\includegraphics[width=0.4\textwidth,height=0.2485\textheight]{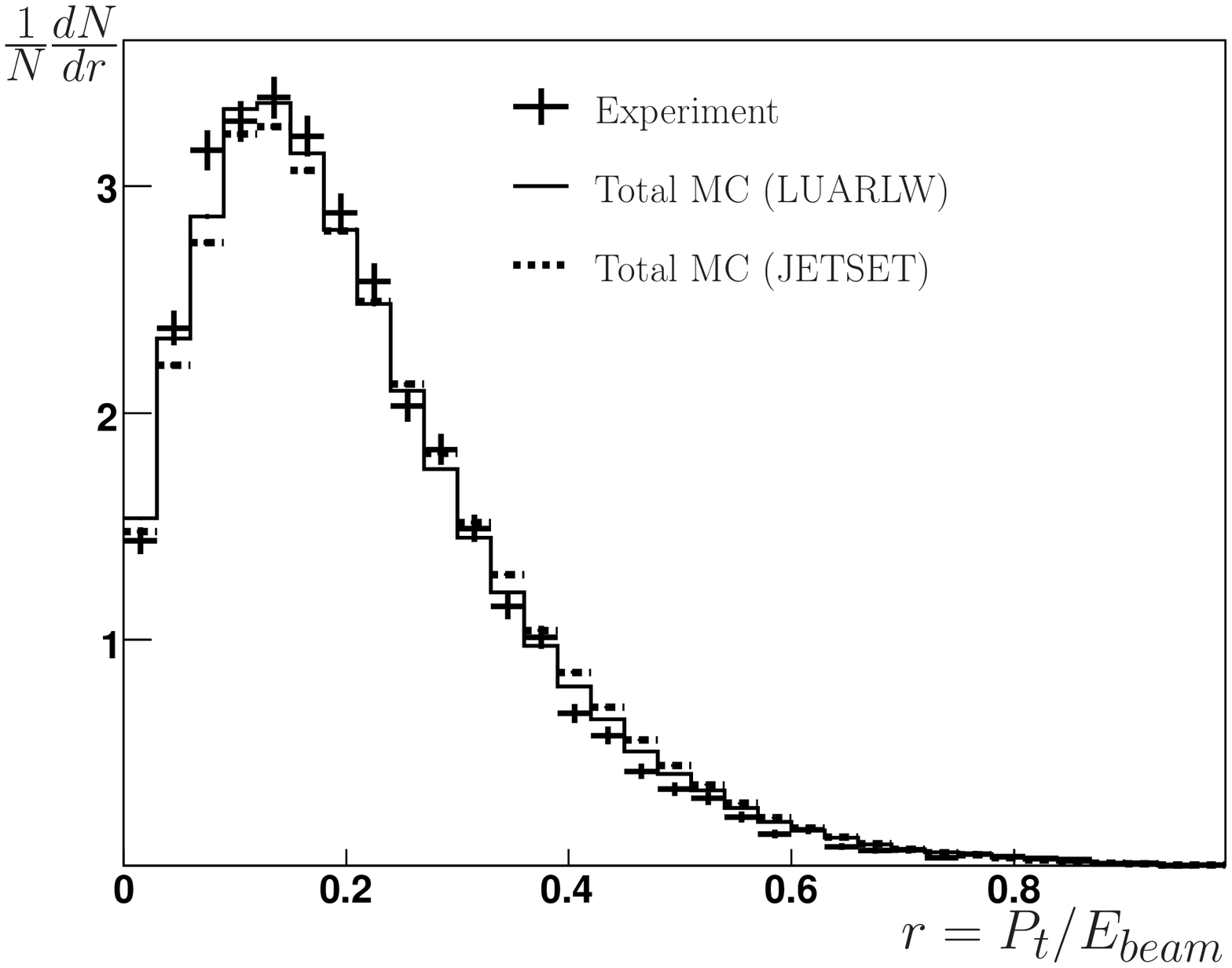}  \\ [1pc]
\includegraphics[width=0.4\textwidth,height=0.25\textheight]{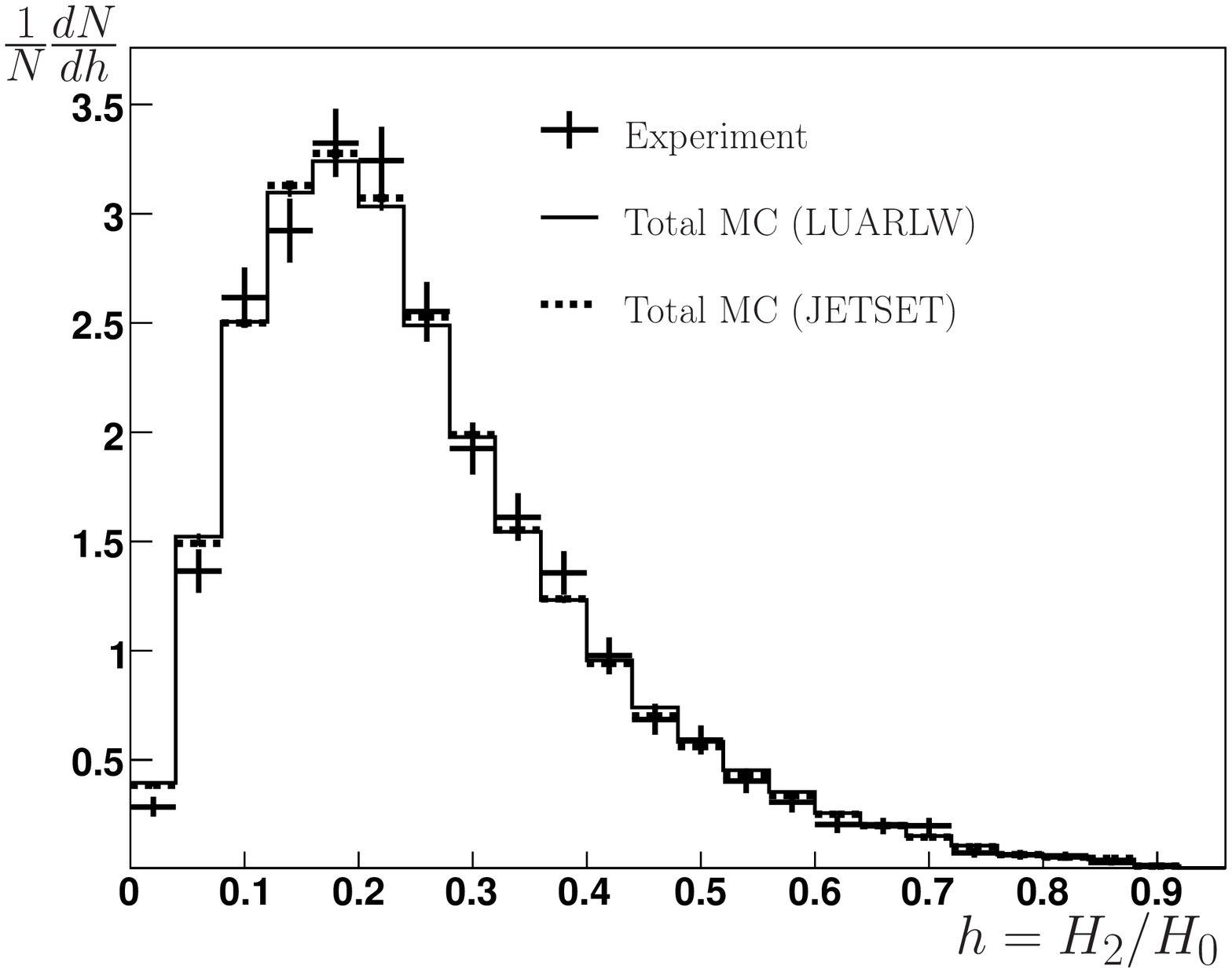}
\hspace*{0.2cm}
\includegraphics[width=0.4\textwidth,height=0.251\textheight]{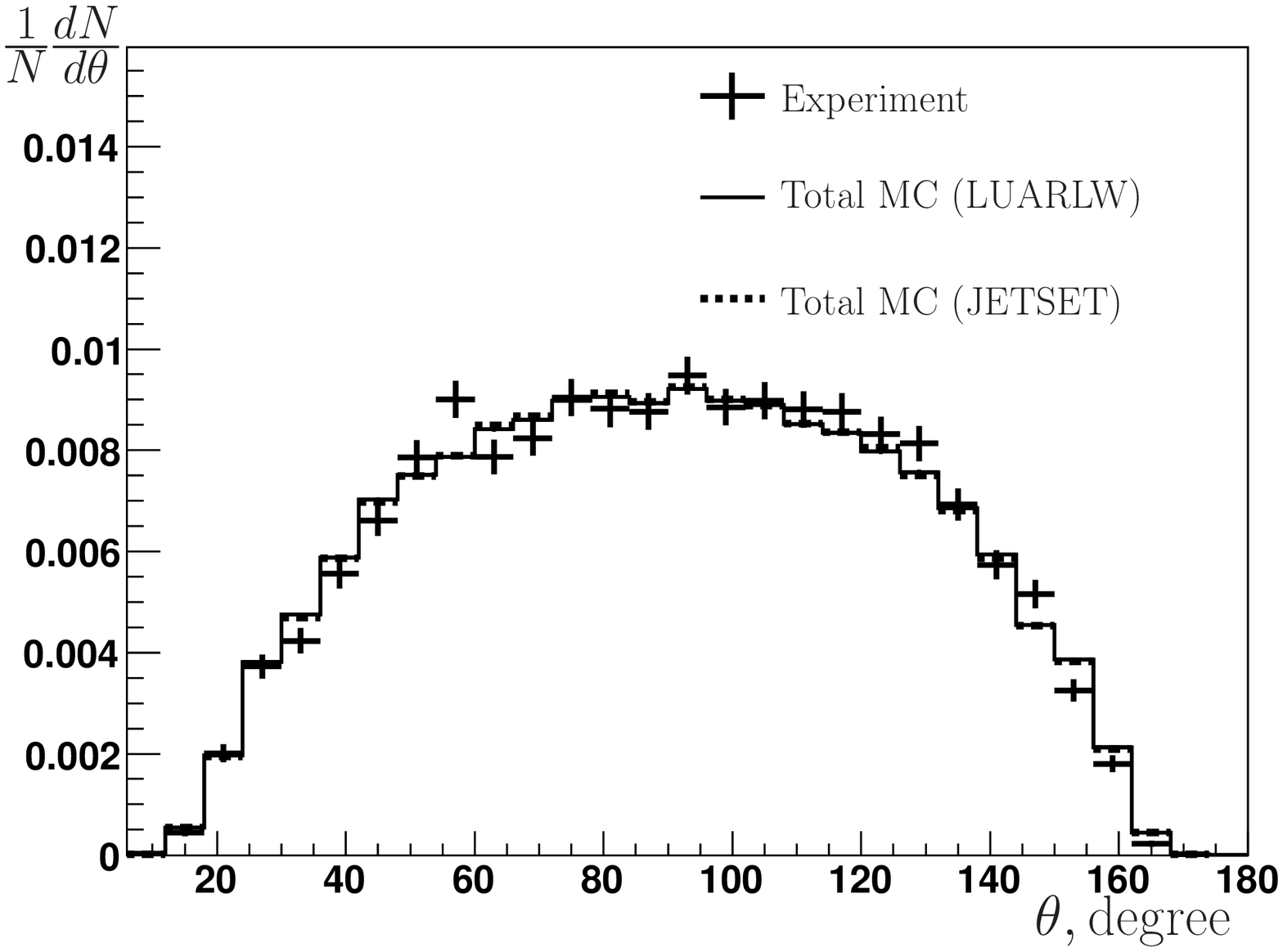}
\\ [1pc]
\includegraphics[width=0.4\textwidth,height=0.25\textheight]{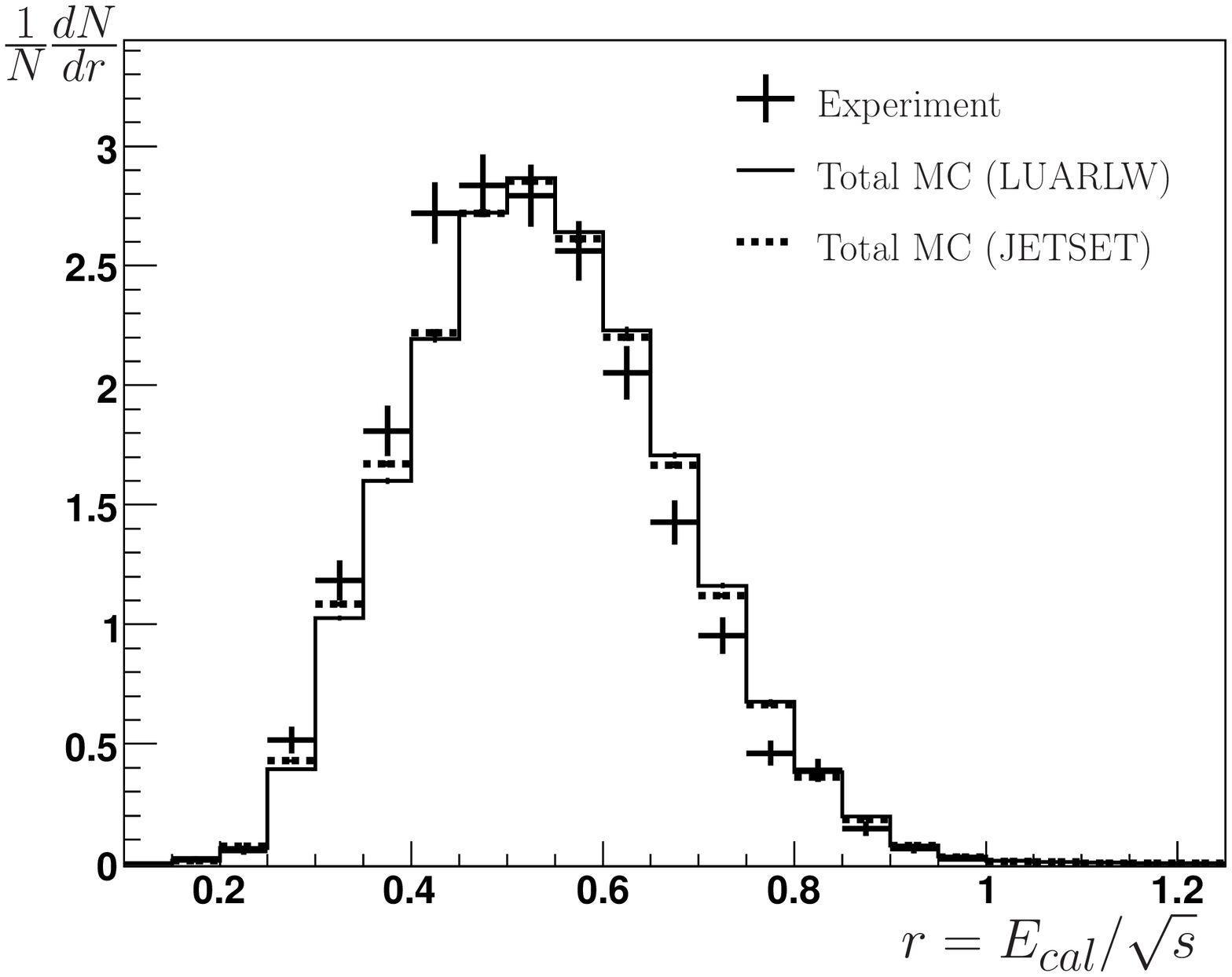}
\hspace*{0.2cm}
\includegraphics[width=0.4\textwidth,height=0.2515\textheight]{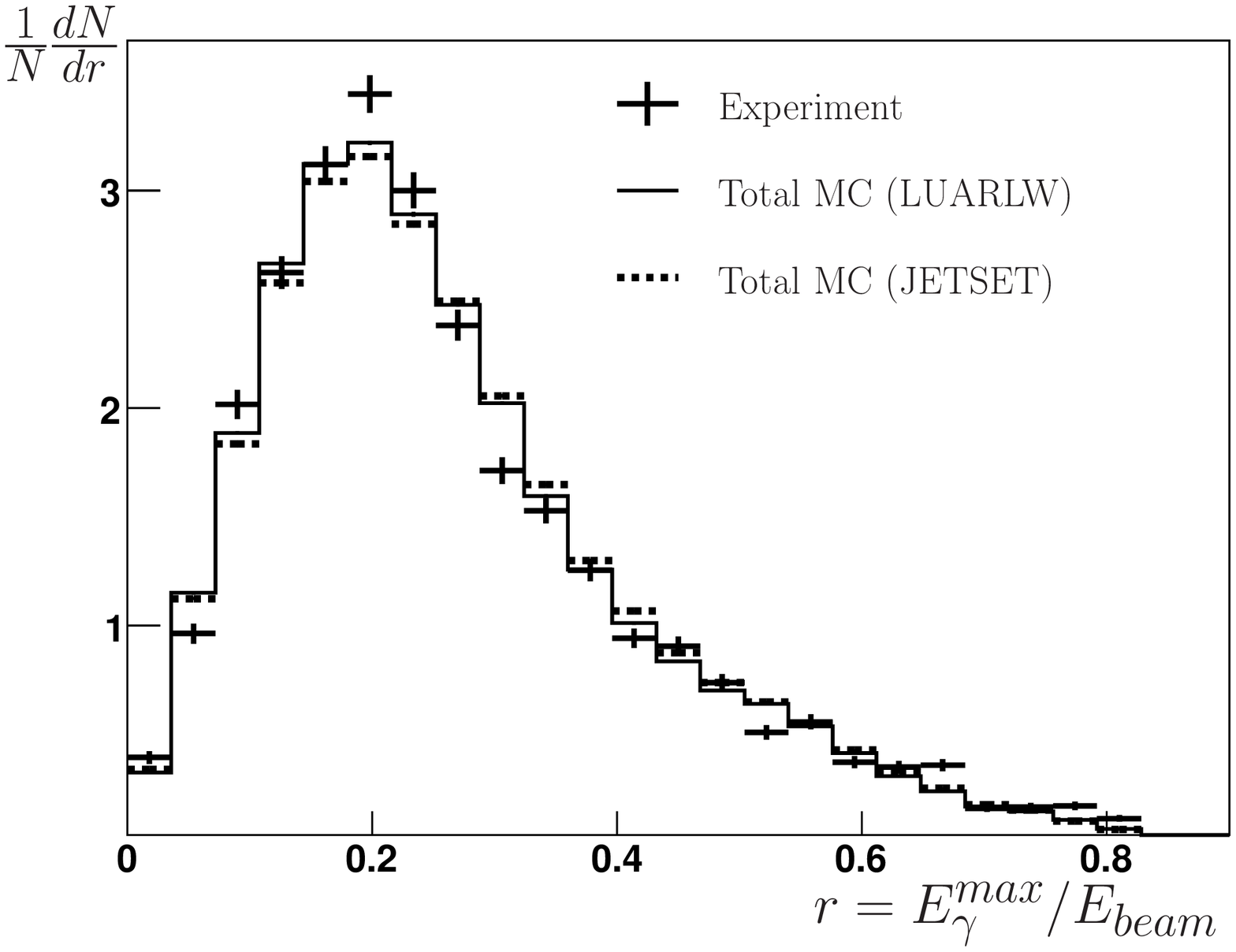} 
\caption{{ 
 Properties of hadronic events produced in the $\text{uds}$ continuum at
 3.119 GeV. Here $N$ is the number of events,
 $N^{IP}_{trk}$ is the number of tracks originating from the interaction 
 region, $P_{t}$ is a transverse momentum of the track,
 $H_2$ and $H_0$ are the Fox-Wolfram moments \cite{Fox:Wolfram},
 $\theta$ is a polar angle of the track,
 $E_{\text{cal}}$ is  energy deposited in the calorimeter, 
 $E_{\gamma}^{\text{max}}$ is  energy of the most energetic photon. 
 The experimental distribution and two variants of MC simulation based on 
 LUARLW and JETSET are plotted. 
 Total MC includes simulation of the $\text{uds}$ continuum, contributions 
of the narrow resonances and leptonic channels, 
we also added the contribution of residual machine background obtained 
from experimental runs with separated electron and positron beams.
 The $P_ {t}$ and polar angle distributions include all tracks in the events.
 The error bars represent statistical errors only.
 All distributions are normalized to unity. 
}}
\label{simdist_fig2}
\end{center}
\end{figure*}
\end{center}

Some important event characteristics are presented in
Fig.~\ref{simdist_fig2}, from which one can see that the experimental and 
simulated distributions agree rather well.  

\subsection{Event selection and detection efficiencies}\protect\label{subsec:mhsel}

In the offline analysis  both experimental and simulated events pass the 
software event filter. By using a digitized response of the detector subsystems
the software filter recalculates the PT and ST decisions 
with stringent conditions. This procedure reduces a systematic uncertainty 
due to trigger instabilities  and
uncertainties on the hardware thresholds.

To suppress the machine background to an acceptable level, the 
following PT conditions were used by OR:
\begin{itemize}\itemsep=-2pt
\item signals from $\ge$ two  non-adjacent scintillation counters\,,
\item signal from the LKr calorimeter\,, 
\item coincidence of the signals from two CsI endcaps.
\end{itemize}
Signals from two particles with the angular separation $\gtrsim\!20^\circ$
should satisfy numerous ST conditions.

The MC simulation shows that the trigger efficiency 
for continuum $\text{uds}$ production
increases from $96.2\%$  at 3.08 GeV to $98.0\%$ at 3.72 GeV.
\renewcommand{\arraystretch}{1.2}
\begin{table}[ht!]
\caption{ \label{tab:criteria} {  Selection criteria for 
hadronic events which were used by AND.}}
\begin{center}
\begin{tabular}{l|l}                                                    
Variable                                      &    Allowed range \\ \hline   
\multicolumn{2}{c}{ $N_{\text{particles}} \geq 3~\text{OR}~\tilde{N}^{\text{IP}}_{\text{track}} \geq 2 $}\\\hline              
$N^{\text{IP}}_{\text{track}}$                &    $\geq 1$\\ \hline        
$E_{\text{obs}}$                              &    $>1.6~\text{GeV}$ \\ \hline
$E_{\gamma}^{\text{max}}/E_{\text{beam}}$     &    $<0.82$ \\ \hline
$E_{\text{cal}}$                 &    $>0.65~\text{GeV}$ \\ \hline
$H_2/H_0$                                     &    $<0.9$ \\ \hline
$|P_{\text{z}}^{\text{miss}}/E_{\text{obs}}|$ &    $<0.6$ \\ \hline
$E_{\text{LKr}}/E_{\text{cal}}$  &    $>0.15$ \\ \hline
$|Z_{\text{vertex}}|$                    &    $<15.0~\text{cm}$\\ \hline                                    
\end{tabular}
\end{center}
\end{table}
\renewcommand{\arraystretch}{1.1}
\begin{table}[t!]
\caption{ \label{tab:def} {Detection efficiency for the $\text{uds}$ continuum} in $\%$ (statistical errors only).}
\begin{center}
\begin{tabular}{cccc}    
Point               &  $\epsilon_{JETSET}$      &  $\epsilon_{LUARLW}    $   &   $\delta \epsilon/\epsilon$    \\ \hline                  
1           & $76.91 \pm  0.13 $   & $76.77 \pm  0.13 $& $-0.2\pm  0.2$   \\ \hline   
2           & $76.77 \pm  0.13 $   & $76.95 \pm  0.13 $& $+0.2\pm  0.2$  \\ \hline   
3           & $77.09 \pm  0.13 $   & $76.96 \pm  0.13 $& $-0.2\pm  0.2$   \\ \hline  
4           & $79.22 \pm  0.13 $   & $80.11 \pm  0.13 $& $-1.1\pm  0.2$   \\ \hline  
5           & $80.38 \pm  0.13 $   & $80.34 \pm  0.13 $& $-0.0\pm  0.2$   \\ \hline  
6           & $80.47 \pm  0.13 $   & $79.98 \pm  0.13 $& $-0.6\pm  0.2$   \\ \hline  
7           & $80.56 \pm  0.13 $   & $80.73 \pm  0.13 $& $+0.2\pm  0.2$   \\ \hline  
8           & $84.03 \pm  0.12 $   & $83.84 \pm  0.12 $& $-0.2\pm  0.2$    \\ \hline 
\end{tabular}
\end{center}
\end{table}
\renewcommand{\arraystretch}{1.1}

Selection criteria for multihadronic events  are presented in 
Table~\ref{tab:criteria} and their description is given below. Here 
$N^{IP}_{track}$ is the number 
of tracks originated from the interaction 
region defined as: 
\mbox{$\,\rho\!<\!5$}~mm,\, \mbox{$|\text{z}_0|\!<\!130$}~mm, 
where $\rho$ is the track impact parameter relative to the beam axis
and  $\text{z}_0$ -- the coordinate of the closest approach point.
The $\tilde{N}^{IP}_{track}$ is the number of tracks satisfying the 
conditions above with $E/p$  less than 0.6, where $E/p$ means the ratio of 
the energy deposited in the calorimeter to the measured momentum of the 
charged particle.
The multiplicity $N_{\text{particles}}$ is a sum of the number of charged tracks 
and the number of neutral particles detected in the calorimeters.

The observable energy  $E_{\text{obs}}$ is defined as a sum  
of the photon energies measured in the electromagnetic calorimeter  and
charged particle energies computed from the track momenta by 
assuming pion masses.
The observable energy cut and limitation on the ratio of the energy of 
the most energetic photon to the beam energy 
$E_{\gamma}^{\text{max}}/E_{\text{beam}}$ suppress 
hadronic events produced via initial-state radiation and thus reduce the 
uncertainty of radiative corrections.   
The total calorimeter energy $E_{\text{cal}} $ is defined as a sum 
of the energies of all clusters 
in the electromagnetic calorimeter.  The requirement on it suppresses 
background from cosmic rays whereas the condition on the ratio of 
the Fox-Wolfram moments $H_{2}/H_{0}$~\cite{Fox:Wolfram} is efficient 
for suppression of the $\ee\!\to\!l^+l^-(\gamma)$ ($l=e,\mu,\tau$)
background, that of cosmic rays and some kinds of the machine background.
The background from two-photon and beam-gas events is suppressed by the 
requirement on the ratio 
$|P_{\text{z}}^{\text{miss}}/E_{\text{obs}}|$,  where $P_{\text{z}}^{\text{miss}}$ 
is the $\text{z}$ component of the missing momentum.
The background from beam-gas events was also suppressed by the condition 
on the ratio  $E_{\text{LKr}}/E_{\text{cal}}$ 
of the energy deposited in the LKr calorimeter and total calorimeter energy.
The event vertex position  $Z_{\text{vertex}}$ is the average
of the $\text{z}_0$'s of the charged tracks. The condition on 
$|Z_{\text{vertex}}|$ suppresses the
background due to beam-gas, beam-wall and cosmic rays.

In addition, the  cosmic background is suppressed with the 
time-of-flight condition and the muon system
veto in the cases when more than two tracks did not cross the interaction 
region.

By applying the selection criteria for hadronic events described above,
we determined the detection efficiencies
for eight data points at which the quantity $R_{\text{uds}}$~was measured.
These values were obtained by using two versions of event simulation
and are listed in Table~\ref{tab:def}.
The detection efficiency at point 8 increased drastically mainly due
to repairing a significant number of calorimeter channels.

\subsection{Luminosity determination}\protect\label{subsec:lum}
The integrated luminosity at each point was
determined by using Bhabha events detected in the LKr calorimeter in the
polar angle range $44^{\circ}\!<\!\theta\!<\!136^{\circ}$. 
The criteria for \ee event selection are listed below:
\begin{itemize}\itemsep=-2pt
\item two clusters, each with the energy above $20\%$ of the beam 
      energy;
\item acollinearities of the polar $\delta \theta$ and azimuthal $\delta \phi$ 
angles are less than $18^{\circ}$;
\item the total energy of these two clusters exceeds
      the single beam energy;
\item the calorimeter energy not associated  with
      these two clusters does not exceed 20$\%$ of the total one;
\item  the ratio of the Fox-Wolfram moments $H_2/H_0 > 0.6$. 
\end{itemize}

To reject the background from 
$\ee\!\to\!\gamma\gamma,\ee\!\to\!e^+e^-e^+e^-$ and
$\ee\!\to\!\text{\it hadrons}$ at least one but not more than three tracks 
originating from the
interaction region were required.

\subsection{Background processes}\protect\label{sec:physbackground}
To determine the $R_{\text{uds}}$ values, we took into account the 
lepton pair production from the QED processes $\ee\to\ee$, $\ee\to\mumu$ and
$\ee\to\tau^{+}\tau^{-}$ which are summarized in Table \ref{tab:effbckg}.

The contributions of two-photon interactions were studied based on the
simulation of $\ee\to\ee X$~events. 
We  found that the contribution of two-photon 
events to the continuum cross section grows from $0.47\%$ at 3.077 GeV 
to $0.51\%$ at 3.72 GeV.
The estimated uncertainty in the $R_{\text{uds}}$ value due to this contribution
is less than $0.2\%$.

\begin{table}[ht]
\caption{\label{tab:effbckg} The contribution of the lepton pair production
to the observed cross section in \%.}
\begin{center}
\begin{tabular}{lccc}        
Point           &\multicolumn{3}{c}{ Process}  \\ 
                & $\ee$          &   $\mumu$       &  $\tau^+\tau^-$             \\ \hline
 1              & $5.06 \pm 0.24$ &  $1.29 \pm 0.27$  &    \\ \hline 
 2              & $1.67 \pm 0.09$ &  $0.42 \pm 0.12$  &    \\ \hline 
 3              & $3.34 \pm 0.17$ &  $0.72 \pm 0.19$  &    \\ \hline 
 4              & $4.03 \pm 0.19$ &  $0.72 \pm 0.15$  &     \\ \hline
 5              & $4.01 \pm 0.20$ &  $0.69 \pm 0.16$  &    \\ \hline  
 6              & $3.42 \pm 0.19$ &  $0.49 \pm 0.16$  &        \\ \hline                
 7              & $4.14 \pm 0.21$ &  $0.53 \pm 0.15$  & $3.37\pm 0.17$ \\ \hline 
 8              & $2.34 \pm 0.13$ &  $0.33 \pm 0.11$  & $4.05\pm 0.20$ \\ \hline 
\end{tabular}
\end{center}
\end{table}

\subsection{Correction for residual machine background}
\protect\label{sec:background}
Our estimates of the contributions of the residual machine background to
the observed hadronic cross section
at different energy points are  listed in the column marked "Method 1" 
of Table~\ref{tab:background}.
These values were obtained by using runs with separated $e^{+}$ and
$e^{-}$ bunches, which were recorded at each energy point.

The number of events which passed selection criteria in the background
runs was used to evaluate the residual background under 
the assumption that the background rate is proportional to the beam
current and the measured vacuum pressure.

As a cross check, we assumed that the background rate is proportional 
to the current only. The results are presented in the last column 
of Table~\ref{tab:background}, which is  marked as "Method 2". 
The maximal difference of 0.28$\%$ between the numbers 
of background events obtained with these two alternatives was
considered as an estimate of the corresponding systematic uncertainty.

\begin{table}[h!]
\caption{{\label{tab:background} The residual machine 
background in $\%$ of the observed cross section}}
\begin{center}
\begin{tabular}{ccc}
Point   &  \multicolumn{2}{c}{ Background in $\%$ (statistical errors only).}   \\\hline        
        &   Method 1 &  Method 2  \\\hline        
    1   &   $1.35 \pm 0.27$&  $1.29 \pm 0.27$\\ \hline   
    2   &   $0.65\pm 0.14$& $0.80\pm 0.15$\\ \hline   
    3   &   $0.81\pm 0.20$& $0.86\pm 0.21$ \\ \hline   
    4   &   $3.80\pm 0.35  $ &  $4.08\pm 0.36$\\ \hline   
    5   &   $2.33\pm 0.30  $ & $2.19\pm 0.29 $ \\ \hline   
    6   &   $1.09\pm 0.23  $& $1.15\pm 0.24$\\ \hline   
    7   &   $0.75 \pm 0.17  $& $0.76\pm 0.18$   \\ \hline   
    8   &   $1.82\pm 0.25  $& $1.94 \pm 0.26$ \\ \hline         
\end{tabular}
\end{center}
\end{table}
\subsection{Radiative correction}
\protect\label{sec:radeff}
Numerical calculation of the radiative correction factor 
was performed  according to  Eq.~\eqref{eq:RadDelta} by 
using the compilation of the vacuum polarization data by the CMD-2 group
 \cite{Actis:2010gg} and the relation between
$R(s)$ and the hadronic part of the vacuum polarization 
$\Pi_{\text{hadr}}(s)$:
\begin{equation}
R(s)=-\frac{3}{\alpha} \Imag \Pi_{\text{hadr}}(s).
\end{equation}

To obtain the quantity $\tilde{R}$ and 
the operator $\tilde{\Pi}$ for Eq.~\eqref{eq:RadDelta}
the contribution of the \JP and \PP 
was subtracted analytically from the vacuum polarization data.

The $\text{uds}$ continuum below 3.077~GeV was simulated 
with the $\text{LUARLW}$ generator, that allows us to determine 
the detection efficiency versus the energy radiated in the initial state.

\begin{figure}[!ht]
\includegraphics[width=0.48\textwidth]{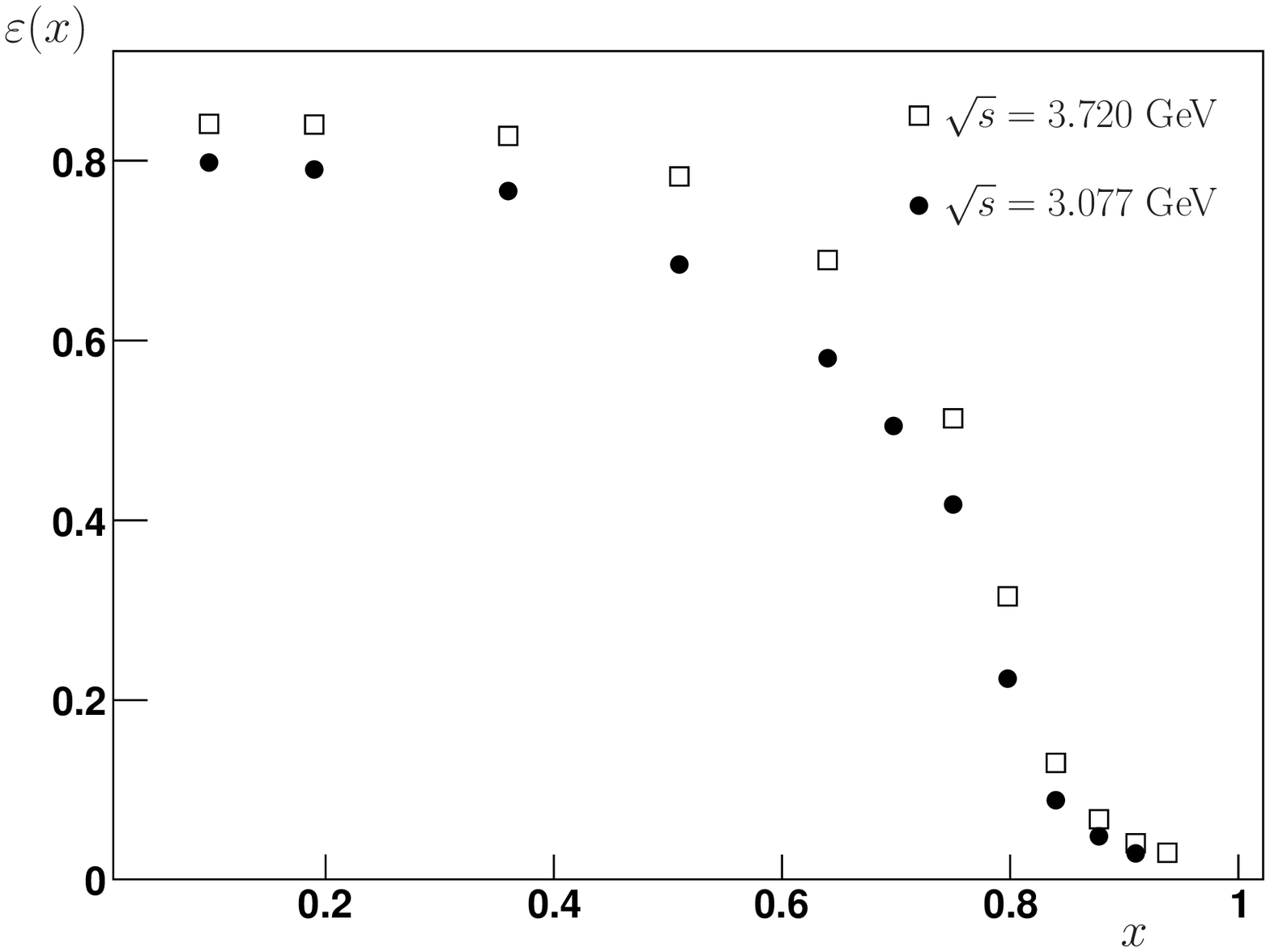}
\caption{{\label{fig:radeff} Hadronic detection efficiency versus
  the  variable $x$ of Eq.~\eqref{eq:RadDelta} at  3.077 GeV and  3.72 GeV.
}}
\end{figure}

The $x$ dependence of the detection efficiency  is shown in 
Fig.~\ref{fig:radeff}. 

The radiative correction factors  at different center-of-mass energies 
are listed in Table \ref{tab:delta}, while the presented systematic 
uncertainties will be discussed in more detail in Sec.~\ref{sec:raderr}.
\begin{table}[th!]
\caption{ \label{tab:delta}{Radiative correction factor $1+\delta$.}}
\begin{center}
\begin{tabular}{ccc}                                           
 Point              &     $1+\delta$      \\ \hline
1              & $1.1091 \pm 0.0089 $    \\ \hline
2              & $1.1108 \pm 0.0089 $      \\ \hline
3              & $1.1120 \pm 0.0056 $      \\ \hline
4              & $1.1130 \pm 0.0078 $      \\ \hline
5              & $1.1133 \pm 0.0067 $     \\ \hline
6              & $1.1151 \pm 0.0056 $      \\ \hline
7              & $1.1139 \pm 0.0078 $      \\ \hline
8              & $1.1137 \pm 0.0056 $      \\ \hline
\end{tabular}
\end{center}
\end{table}

\subsection{$J/\psi$ and $\psi(2S)$ contributions}\protect\label{sec:Jpsi}
To determine contributions of narrow resonances to the observed
cross section we applied resonance parameters retrieved from the fits.
The values presented in Table~\ref{tab:fits}  were corrected for 
the presence of ISR photons.
The corrections were obtained via simulation of $J/\psi$ and
$\psi(2S)$ hadronic decays at each energy point. 

The detection efficiencies obtained from simulation of hadronic decays
in vicinity of narrow resonances are  $(79.00 \pm 0.06) \%$ and $(81.40
\pm 0.08)\%$  for $J/\psi$ and $\psi(2S)$, respectively. 
For both resonances the detection efficiencies obtained by simulation 
agree with the fit results within the estimated errors.

\subsection{Results of energy scan}\protect\label{sec:rscans}
The results of the $R_{\text{uds}}$ measurement obtained in the energy scan 
are presented in Table~\ref{tab:rvalues_sc}.
\begin{table}[ht!]
\caption{\label{tab:rvalues_sc}{ Resulting $R_{\text{uds}}$ values with 
their statistical errors.}}
\begin{center}
\begin{tabular}{lcc}     

     Point      &   $R_{\text{uds}}$  \\ \hline     
 1              & $2.188 \pm 0.056 $  \\ \hline
 2              & $2.211 \pm 0.046 $  \\ \hline   
 3              & $2.214 \pm 0.055 $  \\ \hline   
 4              & $2.233 \pm 0.044 $   \\ \hline  %
 5              & $2.197 \pm 0.047 $   \\ \hline  %
 6              & $2.224 \pm 0.054 $   \\ \hline  %
 7              & $2.220 \pm 0.049 $    \\\hline  %
 8              & $2.213 \pm 0.047 $    \\ \hline %
\end{tabular}
\end{center}
\end{table}

\section{Systematic uncertainties and results}\protect\label{sec:systerr}

\subsection{Systematic uncertainty of absolute luminosity determination}\protect\label{sec:lumerr}
The dominant  contributions to the systematic error of the
absolute luminosity determination with the LKr calorimeter are 
presented in Table~\ref{tab:lumerr}.

\begin{table}[h]
\caption{{ \label{tab:lumerr} Systematic uncertainties of
the luminosity determination.}}
\begin{center}
\begin{tabular}{lc}
Source &  Uncertainty, $\%$   \\ \hline                            
Cross section calculation     & 0.4  \\  \hline 
Calorimeter response          & 0.4  \\  \hline
Calorimeter alignment         & 0.2  \\  \hline
Polar angle resolution        & 0.1  \\  \hline
Background                    & 0.1  \\  \hline
MC statistics                 & 0.1 \\   \hline
Variation of cuts             & 0.7  \\ \hline\hline
Sum in quadrature             & 0.9  \\
\end{tabular}
\end{center}
\end{table}

The uncertainty of the theoretical Bhabha cross section
was evaluated by comparing the results obtained with
the BHWIDE~\cite{BHWIDEGEN} and MCGPJ~\cite{MCGPJ} event generators
at all energy points. The maximum difference did not exceed 0.4$\%$
and agreed with the accuracy quoted by the authors.

The systematic uncertainty related to the imperfect simulation of the 
calorimeter response is about 0.4$\%$. 
It was quantified by variation of relevant simulation parameters such as
the accuracy of the electronic channel calibration, the geometrical
factor controlling sensitivity to the energy loss fluctuations between
calorimeter electrodes, etc.

The alignment of the tracking system and LKr calorimeter is obtained by 
reconstructing cosmic rays.
By using the primary-vertex distribution of multihadronic and Bhabha
events we  determined the interaction point 
position and direction of the beam line.
The luminosity   uncertainty  due to inaccuracy of the alignment  is
about $0.2\%$.

The uncertainty related to the difference of the polar angle resolution
in simulation and data because of event migration into or out of 
the fiducial volume is less than $0.1\%$.

The background to Bhabha events from the 
processes  $\ee \to \mu\mu(\gamma)$ and $\ee \to \gamma\gamma $ and 
$J/\psi$ and $\psi(2S)$ decays contributes less than 0.2\% to the 
observed $e^{+}e^{-}$ cross section
at eight energy points listed in Table~\ref{tab:epoints}.
It was estimated using MC simulation. 
At the complementary points of the scan used for the determination 
of the $J/\psi$ and $\psi(2S)$ parameters the contributions of the 
resonance decays to $e^{+}e^{-}$  were calculated by the fitting.

The luminosity uncertainty due to the residual machine background does not
exceed $0.1\%$.

In addition, we varied requirements within the fiducial region to evaluate
the effect of other possible sources of a systematic uncertainty.
The conditions on the polar angle were varied in a range much larger than the
angular resolution, the variation in the Bhabha event count reaches $50\%$. 
The requirement on the deposited energy was varied in the range of $70-90\%$
of the c.m. energy.  The sum in quadrature of all errors obtained by 
variation of the selection criteria is about $0.7\%$ and gives  
an additional estimate of the systematic uncertainty.
Despite possible double counting we add this 
error to the total luminosity uncertainty to obtain 
a conservative error estimate.

\subsection{Uncertainty due to imperfect simulation of continuum}\protect\label{sec:mchadrerr}
The systematic uncertainty in the $R_{\text{uds}}$ value associated with 
imperfect simulation of the $\text{uds}$ continuum 
was evaluated by using two different MC simulation models. 
We considered the detection efficiencies at eight energy points reported 
in Table~\ref{tab:def} obtained with the JETSET 
and LUARLW hadronic generators. 
It does not exceed a value of 1.1$\%$ which was taken as the systematic 
uncertainty related to the detection efficiency. This estimate is consistent 
with  our previous result of $1.3\%$ obtained in Ref.~\cite{KEDR:R2016} and 
agrees with a value of $0.6\%$ found by the
variation of selection criteria in Sec.~\ref{err:deterr}

\subsection{Systematic uncertainty of the radiative correction}\protect\label{sec:raderr}
The major sources of systematic uncertainty in the radiative correction factor 
at each energy point are presented in Table \ref{tab:raderr}.

To evaluate the uncertainty related to a choice of the vacuum
polarization operator, two alternatives  are compared.
The first one was taken from the CMD-2 work~\cite{Actis:2010gg},
the second was obtained from the BES event generator~\cite{BESGEN}.
The difference in the results obtained according to the data of the
used variants reaches 0.8$\%$ at the points closest to $J/\psi$ 
and  varied from $0.1\%$ to $0.5\%$ at the other points.

The contribution denoted as $\delta R_{\text{uds}}(s)$ is associated with the 
$R_{\text{uds}}(s)$ uncertainty. It is less than 0.2$\%$ for the entire  
energy range.
The contribution $\delta \epsilon(s)$  of about 0.4\% is related 
to the uncertainty in the $\epsilon(s)$ dependence.
A calculation of the radiative corrections according to
Eq.~\eqref{eq:RadDelta} requires the interpolation of the detection 
efficiency presented in Fig.~\ref{fig:radeff} as a function of $x$. 
The contribution $\delta_{\rm calc}$ is related to the interpolation 
uncertainty. It was estimated by comparing the results obtained using 
the linear interpolation and the quadratic one.

\begin{table}[h!]
\caption{{ \label{tab:raderr} Systematic uncertainties of the 
radiative correction.}}
{\small
\begin{center}
\begin{tabular}{cccccc}
             &  \multicolumn{5}{c}{ Uncertainty, $\%$ }     \\ \hline
  Point  &  \multicolumn{4}{c}{Contributions} & Total \\ \hline
   & $\Pi$ approx. &  $\delta R_{\text{uds}}(s)$  &$\delta \epsilon(s)$ & $\delta_{calc}$ &  \\ \hline    
    1  &   0.7 & 0.2&0.4 &0.1 &0.8    \\ \hline       
    2  &   0.7 & 0.1&0.4 &0.1 &0.8    \\ \hline    
    3  &   0.2 & 0.1&0.4 &0.1 &0.5    \\ \hline 
    4  &   0.5 & 0.1&0.4 &0.1 &0.7    \\ \hline
    5  &   0.4 & 0.1&0.4 &0.1 &0.6    \\ \hline  
    6  &   0.2 & 0.1&0.4 &0.1 &0.5    \\ \hline 
    7  &   0.5 & 0.1&0.4 &0.1 &0.7    \\ \hline    
    8  &   0.1 & 0.2&0.4 &0.1 &0.5    \\ \hline    
\end{tabular}
\end{center}
}
\end{table}

\subsection{Detector-related uncertainties in $R_{\text{uds}}$}\label{err:deterr}
The track reconstruction efficiency was studied by using Bhabha events and
low-momentum cosmic tracks and the appropriate correction was
introduced in the MC simulation. The uncertainty of the correction
introduces an additional systematic uncertainty of about $0.2\%$. 
We also used two methods to achieve data and MC 
consistency in the momentum and angular resolution.
The first way was to scale the spatial resolution of the drift chamber,
while the second method assumed  scaling  systematic uncertainties 
of the calibration parameters of the tracking system.
The maximal obtained variation of the detection efficiency at
various energies is less than 0.3$\%$.
Thus, the uncertainty related to track reconstruction is  about 0.4\%.

The trigger efficiency uncertainty is about 0.2$\%$ and mainly comes from 
the calorimeter thresholds in the secondary trigger. 
It was estimated by varying the threshold in the software event filter. 

The trigger and  event selection efficiency depend 
on the calorimeter response to hadrons. 
We estimated the uncertainty of 0.2$\%$ related to the simulation of 
nuclear interaction by comparing the efficiencies obtained with the
packages GHEISHA~\cite{Fesefeldt:1985yw} and FLUKA~\cite{Fasso:2005zz}
which are implemented in GEANT~3.21~\cite{GEANT:Tool}. 

The effect of other possible sources of the detector-related uncertainty 
was evaluated by varying the event selection criteria that are presented 
in Table~\ref{tab:criteriavar}. 
Tightening of some requirements listed in Table~\ref{tab:criteriavar}
by several times varies a contribution to the observed cross section of
physical and machine background events and significantly changes the
detection efficiency. That allows us to verify uncertainties associated 
with the background and radiative corrections.

All observed $R_{\text{uds}}$ variations were smaller than their statistical 
errors and 
can originate from the already considered sources of uncertainties or  
statistical fluctuations.
By keeping a conservative estimate,  we added them in the total uncertainty.

\begin{table}[h!]
\caption{\label{tab:criteriavar} { $R_{\text{uds}}$ uncertainty due to 
variation of the selection criteria for hadronic events.}}
\begin{center}
\begin{tabular}{llc}      
Condition$/$Variable                                      &   Range  variation & $R_{\text{uds}}$ variation in \% \\
                                    
$N_{\text{particles}} \geq 3~\text{OR}~$&$N_{\text{particles}} \geq 4~\text{OR}~$ &0.1\\ 
$\tilde{N}^{\text{IP}}_{\text{track}} \geq 2 $&$\tilde{N}^{\text{IP}}_{\text{track}} \geq 2 $ \\ \hline         
$N^{\text{IP}}_{\text{track}}$                 &    $\geq 1$ OR no cut  &$0.1$   \\ \hline        
$E_{\text{obs}}$                               &    $> 1.4  \div 1.8~\text{GeV}$ & 0.3 \\ \hline
$E_{\gamma}^{\text{max}}/E_{\text{beam}}$      &    $< 0.6  \div 0.9 $ & 0.3 \\ \hline
$E_{\text{cal}}$                  &    $> 0.5 \div 0.75~\text{GeV}$ & 0.2 \\ \hline
$H_2/H_0$                                      &    $< 0.7  \div 0.93$ &  0.2\\ \hline
$|P_{\text{z}}^{\text{miss}}/E_{\text{obs}}|$  &    $< 0.6  \div 0.8$  & 0.2     \\ \hline
$E_{\text{LKr}}/E_{\text{cal}}$   &    $> 0.15 \div 0.25$  & 0.1\\ \hline
$|Z_{\text{vertex}}|$                          &    $< 20.0 \div  13.0~\text{cm}$&  0.2  \\  \hline\hline 
\multicolumn{2}{c}{Sum in quadrature}          &    0.6 \\                             
\end{tabular}
\end{center}
\end{table}

\subsection{Summary of systematic uncertainties and results}
\protect\label{sec:errsummary}

The major sources of the systematic uncertainty on the $R_{\text{uds}}$ value 
are listed in Table~\ref{tab:rerr}.
\renewcommand{\arraystretch}{1.2}
\begin{table}[h!]
\caption{\label{tab:rerr} $R_{\text{uds}}$ systematic uncertainties in $\%$
assigned to each energy point.}
\begin{center}
\begin{tabular}{lcccccccc}
                         &   1&   2 & 3  &  4 &  5 & 6  &  7 &  8  \\
                         \hline 
Luminosity               & 0.9&  0.9& 0.9& 0.9& 0.9& 0.9& 0.9& 0.9 \\\hline %
Radiative correction     & 0.8&  0.8& 0.5& 0.7& 0.6& 0.5& 0.7& 0.5 \\\hline  
Continuum simulation     & 1.1&  1.1& 1.1& 1.1& 1.1& 1.1& 1.1& 1.1 \\ \hline %
Track reconstruction     & 0.4&  0.4& 0.4& 0.4& 0.4& 0.4& 0.4& 0.4 \\ \hline %
$e^+e^-X$  contribution  & 0.2&  0.2& 0.2& 0.2& 0.2& 0.2& 0.2& 0.2 \\ \hline %
$l^+l^-$  contribution   & 0.4&  0.4& 0.4& 0.3& 0.3& 0.3& 0.4& 0.4 \\ \hline %
Trigger efficiency       & 0.2&  0.2& 0.2& 0.2& 0.2& 0.2& 0.2& 0.2 \\ \hline %
Nuclear interaction      & 0.2&  0.2& 0.2& 0.2& 0.2& 0.2& 0.2& 0.2 \\ \hline %
Cuts variation           & 0.6&  0.6& 0.6& 0.6& 0.6& 0.6& 0.6& 0.6  \\ \hline%
$J/\psi$ and  $\psi(2S)$ & 0.1&  1.8& 0.4& 0.2& 0.1& 0.1& 0.1& 1.1 \\ \hline %
Machine background       & 0.4&  0.8& 0.5& 0.6& 0.5& 0.4& 0.4& 0.6 \\\hline\hline   %
Sum in quadrature        & 1.9&  2.7& 1.9& 1.9& 1.8& 1.8& 1.9& 2.2  \\ %
\end{tabular}
\end{center}
\end{table}

During data collection at a given energy point the relative 
beam energy variation was less than $10^{-3}$  allowing us to 
neglect this source of uncertainty. 

The results obtained at most points supplement the data published in
Ref.~\cite{KEDR:R2016}.
In order to use these data in the calculations of the dispersion integrals
it is important to combine results of both experiments by
taking into account correlated uncertainties properly.
This requires to determine the common components of the uncertainties
which are the same for each experiment. 
The corresponding contributions to the systematic  uncertainty are 
listed in Table~\ref{tab:correlated}.
\begin{table}[h!]
\caption{\label{tab:correlated}Correlated systematic uncertainties $R_{\text{uds}}$ 
in $\%$ for  data of 2011 and 2014.}
\begin{center}
\begin{tabular}{lc}                                        
Source                         &  Uncertainty in $\%$ \\\hline %
Luminosity&          \\ \hline%
 Cross section calculation &  0.4    \\\hline \hline 
Radiative correction&             \\\hline 
 $\Pi$ approx.            &  $0.1 \div 0.3$    \\
 $\delta R_{\text{uds}}(s)$            &  $0.1 \div 0.2$    \\
 $\delta \epsilon(s)$     &  $0.2$ \\\hline \hline 
Continuum simulation           & 0.9   \\ \hline %
$e^+e^-X$  contribution         & 0.1   \\ \hline %
$l^+l^-$  contribution          & 0.2   \\ \hline %
Trigger efficiency             & 0.2   \\ \hline %
Nuclear interaction            & 0.2   \\ \hline\hline  %
Sum in quadrature              & $1.1$   \\ %
\end{tabular}
\end{center}
\end{table}

The results of the two scans were averaged using 
their statistical uncertainties and the uncorrelated parts of the systematic
ones. The formal description of the averaging procedure can be found
in Ref.~\cite{psi2S:2012}.
The obtained $R_{\text{uds}}$ and $R$ values as well as luminosity-weighted 
average center-of-mass energies are presented in Table~\ref{tab:rvalues}. 
As mentioned above, the contribution of narrow resonances 
to $R(s)$ is not negligible in the resonance region.
This contribution was determined  analytically by using "bare"  parameters 
of the resonances, which were calculated based on the PDG data~\cite{PDG:2014}.
The inaccuracy of $R$ associated with the resonance parameters is negligible 
in comparison with the other uncertainties, 
so the errors for the values of $R$ and $R_{\text{uds}}$ are the same.
\begin{table*}[th!]
\caption{\label{tab:rvalues}{Measured values of $R_{\text{uds}}(s)$ and $R(s)$
with statistical and systematic uncertainties.}}
\begin{center}
{
\begin{tabular}{cc|cc|cc}     
\multicolumn{2}{c}{Data 2011 \cite{KEDR:R2016}}       &\multicolumn{2}{c}{Data 2014}&  \multicolumn{2}{c}{Combination} \\
 $\sqrt{s}$, MeV&$R_{\text{uds}}(s)$ &$\sqrt{s}$, MeV&
 $R_{\text{uds}}(s)$ &   $\sqrt{s}$, MeV   & $R_{\text{uds}}(s) \{ R(s)\}$ \\\hline       
 - & -         & $3076.7 \pm 0.2 $ &  $2.188 \pm 0.056 \pm 0.042$  & $3076.7 \pm 0.2 $&  $2.188 \pm 0.056 \pm 0.042$ \\   \hline 
 $3119.9  \pm 0.2$& $2.215 \pm 0.089 \pm 0.066$ & $3119.2 \pm 0.2$& $2.211 \pm 0.046  \pm 0.060$ & $3119.6\pm 0.4$&  $2.212\{2.235\} \pm 0.042  \pm 0.049$  \\ \hline  
 $3223.0  \pm 0.6$& $2.172 \pm 0.057 \pm 0.045$ & $3221.8 \pm 0.2$& $2.214 \pm 0.055  \pm 0.042$ &$3222.5 \pm 0.8$ & $2.194\{2.195\} \pm 0.040  \pm 0.035$ \\ \hline  
 $3314.7  \pm 0.7$& $2.200 \pm 0.056 \pm 0.043$ & $3314.7 \pm 0.4$& $2.233 \pm 0.044  \pm 0.042$ &$3314.7 \pm 0.6$ & $2.219\{2.219\} \pm 0.035  \pm 0.035$ \\ \hline  
 $3418.2  \pm 0.2$& $2.168 \pm 0.050 \pm 0.042$ & $3418.3 \pm 0.4$& $2.197 \pm 0.047 \pm  0.040$ &$3418.3 \pm 0.3$ & $2.185\{2.185\}\pm 0.032   \pm 0.035$ \\ \hline  
 -  & - & $3499.6 \pm 0.4$& $2.224 \pm 0.054 \pm  0.040$ &  $3499.6 \pm 0.4$ & $2.224\{2.224\} \pm 0.054 \pm  0.040$ \\ \hline  
 $3520.8  \pm 0.4$ & $2.200 \pm 0.050 \pm 0.044$ & -& -&  $3520.8  \pm 0.4$ &  $2.200\{2.201\} \pm 0.050 \pm 0.044$  \\ \hline  
 $3618.2  \pm 1.0$& $2.201 \pm 0.059 \pm 0.044$ & $3618.1 \pm 0.4$& $2.220 \pm 0.049  \pm 0.042$ &$3618.2 \pm 0.7$ &  $2.212\{2.218\} \pm 0.038  \pm 0.035$ \\ \hline  
 $3719.4  \pm 0.7$& $2.187 \pm 0.068 \pm 0.060$ & $3719.6 \pm 0.2$& $2.213 \pm 0.047  \pm 0.049$ &$3719.5 \pm 0.5$ &  $2.204\{2.228\} \pm 0.039  \pm 0.042$ \\ \hline  
\end{tabular}
}
\end{center}
\end{table*}

\section{Results}
\begin{figure}[!h]
\begin{center}
\includegraphics[width=0.49\textwidth]{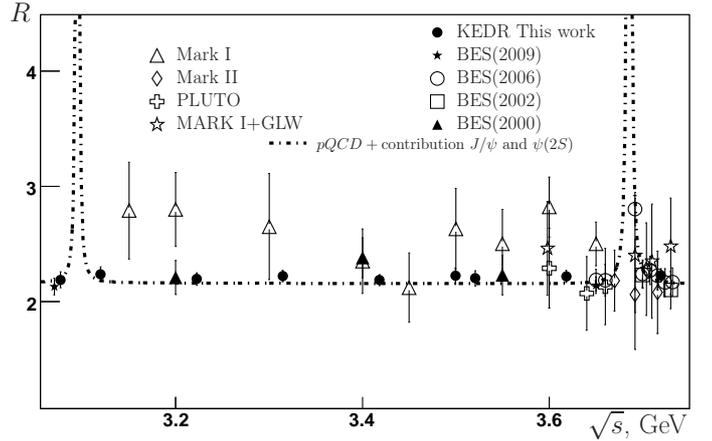}
\end{center}
\caption{{\label{fig:rfinal} The quantity R versus the c.m. energy and 
the sum of the prediction of perturbative QCD and a contribution of 
narrow resonances.
}}
\end{figure}

By combining new data with our previous results we determined
the values of $R_{\text{uds}}$ and $R$ at nine center-of-mass energy points
between 3.08 and 3.72 GeV. The accuracy of $R$ measurements in comparison 
with our previous results ~\cite{KEDR:R2016} was increased by 
$1.4\div 1.7$ times.
The total error is about or better than $2.6\%$ at most
of energy points with a systematic uncertainty of about $1.9\%$. 
This result provides the most precise information about $R$ in this 
energy range. The measured $R$ values are shown in Fig.~\ref{fig:rfinal}. 
For completeness, we remind that in the $R$ measurement performed at
KEDR in the c.m.energy range 1.84 -- 3.05 GeV the total uncertainty was
3.9\% or better with a systematic one of about 2.4\%~\cite{KEDR:R2017}.

In the c.m.energy range 3.08-3.72 GeV the weighted average 
$\overline{R}_{\text{uds}} = 2.204 \pm 0.014 \pm 0.026$ 
is approximately one sigma higher than that theoretically expected, 
$R_{\text{uds}}^{\text{pQCD}}  = 2.16 \pm 0.01$ calculated  according to the pQCD 
expansion~\cite{Baikov:pQCD} 
for  $\alpha_{s}(m_{\tau})=0.333\pm0.013$ obtained from semileptonic $\tau$ 
decays~\cite{Brambilla:2014}. In the lower c.m.energy range 1.84-3.05 GeV
the weighted average is $2.225 \pm 0.020 \pm 0.047$ in good agreement
with the pQCD prediction of $2.18 \pm 0.02$.

It should be noted that while calculating the dispersion integrals in 
this energy range it is preferable to use the measured $R_{\text{uds}}(s)$ values 
by adding the contribution of narrow resonances calculated analytically.
This approach prevents from a possible double counting of the contribution 
of narrow resonances.

\section{Summary and Applications}
Together with the high-precision $R$ measurement below 
the $J/\psi$~\cite{KEDR:R2017}, KEDR measured the $R$ values  
at twenty two  center-of-mass energies 
between 1.84 and 3.72 GeV listed in Table~\ref{summary:table}.
\begin{table}[h!]
\caption{{\label{summary:table} Summary table of KEDR results.  
Actual energies and measured $R$ values.}}
\begin{tabular}{lcc}
Point & Energy          &$R_{\text{uds}}(s) \{ R(s)\}$ \\\hline
\multicolumn{3}{c}{Data 2010 \cite{KEDR:R2017}}  \\\hline
1     &$1841.0 \pm 2$   & $2.226   \pm 0.139 \pm 0.158$     \\\hline
2     &$1937.0 \pm 2$   & $2.141   \pm 0.081 \pm 0.073$    \\\hline
3     &$2037.3 \pm 2$   & $2.238   \pm 0.068 \pm 0.072$     \\\hline 
4     &$2135.7 \pm 2$   & $2.275   \pm 0.072 \pm 0.055$    \\\hline
5     &$2239.2 \pm 2$   & $2.208   \pm 0.069 \pm 0.053$    \\\hline
6     &$2339.5 \pm 2$   & $2.194   \pm 0.064 \pm 0.048$       \\\hline
7     &$2444.1 \pm 2$   & $2.175   \pm 0.067 \pm 0.048$     \\\hline
8     &$2542.6 \pm 2$   & $2.222   \pm 0.070 \pm 0.047$      \\\hline
9     &$2644.8 \pm 2$   & $2.220   \pm 0.069 \pm 0.049$         \\\hline
10    &$2744.6 \pm 2$   & $2.269   \pm 0.065 \pm 0.050$         \\\hline
11    &$2849.7 \pm 2$   & $2.223   \pm 0.065 \pm 0.047$       \\\hline
12    &$2948.9 \pm 2$   & $2.234   \pm 0.064 \pm 0.051$   \\\hline
13    &$3048.1 \pm 2$   & $2.278   \pm 0.075 \pm 0.048$     \\\hline
\multicolumn{3}{c}{Combined Data 2011 \cite{KEDR:R2016} and 2014 (This work)}  \\\hline
14    &$3076.7 \pm 0.2$ &$2.188    \pm 0.056 \pm 0.042$          \\\hline
15    &$3119.6\pm 0.4$  &$2.212\{2.235\} \pm 0.042 \pm 0.049$ \\\hline
16    &$3222.5 \pm 0.8$ &$2.194\{2.195\} \pm 0.040 \pm 0.035$ \\\hline
17    &$3314.7 \pm 0.6$ &$2.219\{2.219\} \pm 0.035 \pm 0.035$ \\\hline
18    &$3418.3 \pm 0.3$ &$2.185\{2.185\} \pm 0.032 \pm 0.035$ \\\hline
19    &$3499.6 \pm 0.4$ &$2.224\{2.224\} \pm 0.054 \pm 0.040$ \\\hline
20    &$3520.8  \pm 0.4$&$2.200\{2.201\} \pm 0.050 \pm 0.044$ \\\hline
21    &$3618.2  \pm 1.0$&$2.212\{2.218\} \pm 0.038 \pm 0.035$ \\\hline
22    &$3719.4  \pm 0.7$&$2.204\{2.228\} \pm 0.039 \pm 0.042$ \\
\end{tabular}
\end{table}

To use $R(s)$ data it is necessary to take into account point-by-point 
correlated effects.
The analysis of the sources of systematic uncertainties makes it
possible to identify common contributions in the listed data sets. 
Similarly to the  Table~\ref{tab:correlated} presented above, the
correlated systematic uncertainties $R_{\text{uds}}$ for other data
sets are listed in Table~\ref{tab:corrdata}. 
\begin{table}[h!]
\caption{{\label{tab:corrdata} Correlated systematic uncertainties of $R_{\text{uds}}$ 
in $\%$ for    data of 2010, 2011 and 2014.}}
\begin{center}
\begin{tabular}{lcc}                                         
Source                         &  \multicolumn{2}{c}{Uncertainty in $\%$}\\\hline %
                               &Data 2010    &  Data 2010 / Data 2011, 2014      \\\hline 
Luminosity                     & \multicolumn{2}{c}{}    \\ \hline%
 Cross section calc.           &  0.5 &  0.4    \\
 Calorimeter response          &  0.7 &  -  \\
 Calorimeter alignment          &  0.2 &  0.2  \\ \hline \hline 
Radiative correction&                     \\\hline 
 $\Pi$ approx.                 & 0.3 & 0.1  \\
 $\delta R_{\text{uds}}(s)$       & 0.2 & 0.2    \\
 $\delta \epsilon(s)$          & 0.3 & 0.2  \\\hline \hline 
Continuum simulation           & 1.2 & $0.4\div 0.8$  \\ \hline %
Track reconstruction           & 0.5 & 0.4  \\ \hline %
$e^+e^-X$  contribution        & 0.2 & 0.1  \\ \hline %
$l^+l^-$  contribution         & 0.3 & 0.2 \\ \hline %
Trigger efficiency             & 0.3 & 0.2 \\ \hline %
Nuclear interaction            & 0.4 & 0.2 \\ \hline\hline  %
Sum in quadrature              & 1.8 &  $0.8\div1.1$   \\ %
\end{tabular}
\end{center}
\end{table}
Keeping a conservative approach, we believe these
contributions are completely correlated, that allows us to write down an 
approximate correlation matrix for
systematic uncertainties (Table~\ref{tab:corrmatrix}). 
Note that statistical errors in our $R$ results are fully
uncorrelated.

The  determination of the $R$ ratio plays a key role in the determination of 
the running strong coupling  constant $\alpha_{s}(s)$.
To verify the compatibility with other measurements of $\alpha_{s}(s)$
we performed  a fit of  $R_{\text{uds}}$ in the given energy range 
using the following approximation \cite{Baikov:pQCD}:
\begin{equation}
R_{\text{uds}}^{\text{calc}}(s)=2\times \bigg (1+\frac{\alpha_{s}}{\pi}+\frac{\alpha^2_{s}}{\pi^2}\times\bigg(\frac{365}{24}-9\zeta_3-\frac{11}{4}\bigg) \bigg)~,
\label{R:exp}
\end{equation}
where $\zeta$ is the Euler-Riemann zeta function and $\alpha_{s}(s)$ is 
approximated by:
\begin{equation}
\begin{split}
\alpha_{s}(s) & =\frac{1}{b_0 t} \Bigg (1-\frac{b_1 l}{b_0^2 t} + \frac{b_1(l^2-l-1)+b_0 b_2}{b^4_0 t^2} \\
+&\frac{b^3_1(-2l^3+5l^2+4l-1)-6 b_0 b_2 b_1 l +  b_0^2 b^3}{2 b^6_0 t^3} \Bigg)~, 
\end{split}
\end{equation}
with $ t=\ln {\frac{s}{\Lambda^2}}$, $l=\ln{t}$ parametrized in terms of the 
QCD scale parameter $\Lambda$  
and coeffients $b_0,b_1,b_3$ defined in \cite{PDG:2018}.

To determine $\Lambda$, we minimise  the $\chi^2$ function
\begin{equation}
\chi^2= \sum_{i}\sum_{j} \bigg(R_{\text{uds}}^{\text{meas}}(s_i)-R_{\text{uds}}^{\text{calc}}(s_i)\bigg) C_{ij}^{-1} \bigg(R_{uds}^{\text{meas}}(s_j)-R_{\text{uds}}^{\text{calc}}(s_j)\bigg)~,
\end{equation}
where $C_{ij}^{-1}$ are coefficients of the inverse covariance matrix which 
is derived 
from statistical errors and systematic uncertainties taking into account 
the correlation matrix presented in  Table~\ref{tab:corrmatrix}.  

The obtained value of  $\Lambda= 0.361^{+0.155}_{-0.174}~\text{GeV}$ 
corresponds to $\alpha_{s}(m_{\tau})=0.332^{+0.100}_{-0.092}$. 
If the next order of pQCD is included in the expansion of $R_{\text{uds}}$, 
the fitting results are as follows: 
$\Lambda= 0.437^{+0.210}_{-0.215}~\text{GeV}$ and 
$\alpha_{s}(m_{\tau})=0.378^{+0.173}_{-0.120}$.
So, we can conclude that our measurements of $R(s)$ are consistent with 
the pQCD predictions within their errors.

Another practical application of the $R(s)$ measurement is related to 
determination of the heavy quark masses.
This calculation is based on sum rules and experimental moments 
$M_n^{\text{exp}}$, which are defined as follows:
\begin{equation}
M_{n}^{\text{exp}}= \int \frac{R(s)}{s^{n+1}} ds ~.
\end{equation}
The inclusion in the analysis of our new results increases the accuracy 
of the contribution of light quarks to experimental moments by almost 
two times in the given energy range.  
According to Ref. \cite{mq:2016},  the total uncertainty of $c$ quark mass 
determination is 8~MeV, 
in which the light quark contribution is about 2~MeV. 
By applying new KEDR results one can reduce this contribution down to 1~MeV.

\begin{table*}[h!]
\caption{{\label{tab:corrmatrix} The correlation matrix for systematic uncertainties of the $R$ values obtained in the KEDR experiments.}}
{\footnotesize
\begin{tabular}{lcccccccccccccccccccccccc}
Point&\multicolumn{22}{c}{Correlation Matrix}  \\  
1 &1 &0.139&0.143&0.193&0.192&0.212&0.212&0.216&0.207&0.211&0.216&0.201&0.222&0.096&0.046&0.096&0.105&0.110&0.098&0.089&0.114&0.071 \\  
2  &  & 1  &0.309&0.418&0.408&0.445&0.437&0.466&0.446&0.457&0.467&0.434&0.480&0.200&0.097&0.201&0.225&0.229&0.212&0.189&0.244&0.151 \\  
3   &   &   & 1 &0.423&0.425&0.470&0.470&0.480&0.460&0.463&0.480&0.442&0.486&0.212&0.101&0.212&0.232&0.243&0.218&0.198&0.253&0.158 \\  
4  &   &   &   & 1&0.575&0.635&0.635&0.649&0.622&0.610&0.649&0.598&0.637&0.287&0.137&0.286&0.314&0.329&0.295&0.268&0.342&0.213 \\  
5 &   &   &   &   & 1&0.621&0.621&0.642&0.615&0.629&0.643&0.598&0.661&0.280&0.134&0.280&0.310&0.322&0.293&0.262&0.336&0.208\\  
6  &   &   &   &   &   & 1&0.677&0.709&0.679&0.695&0.710&0.661&0.730&0.306&0.148&0.305&0.342&0.351&0.323&0.287&0.371&0.229\\  
7  &   &   &   &   &   &   & 1&0.709&0.679&0.695&0.710&0.661&0.730&0.304&0.148&0.305&0.342&0.348&0.323&0.287&0.371&0.229 \\  
8   &   &   &   &   &   &   &   &  1 &0.695&0.710&0.725&0.675&0.745&0.320&0.153&0.320&0.351&0.368&0.330&0.299&0.382&0.238\\  
9  &   &   &   &   &   &   &   &    & 1&0.681&0.695&0.647&0.715&0.307&0.146&0.306&0.336&0.352&0.316&0.287&0.366&0.228\\  
10&   &   &   &   &   &   &   &    &   & 1 &0.710&0.654&0.701&0.314&0.150&0.313&0.344&0.360&0.323&0.293&0.374&0.233 \\  
11 &   &   &   &   &   &   &   &    &   &   & 1 &0.675&0.745&0.321&0.153&0.320&0.351&0.368&0.330&0.300&0.382&0.238\\  
12 &   &   &   &   &   &   &   &    &   &   &   & 1&0.687&0.298&0.142&0.298&0.327&0.342&0.307&0.279&0.356&0.222\\  
13  &   &   &   &   &   &   &   &    &   &   &   &   & 1&0.330&0.157&0.329&0.361&0.378&0.339&0.308&0.393&0.245 \\  
14 &   &   &   &   &   &   &   &    &   &   &   &   &   & 1 &0.288&0.396&0.405&0.394&0.356&0.317&0.403&0.333 \\  
15&    &   &   &   &   &   &   &    &   &   &   &   &   &   & 1 &0.345&0.347&0.345&0.305&0.275&0.345&0.288\\  
16 &    &   &   &   &   &   &   &    &   &   &   &   &   &   &   & 1&0.486&0.475&0.427&0.38&0.483&0.400\\  
17 &   &   &   &   &   &   &   &    &   &   &   &   &   &   &   &   &1&0.486&0.427&0.387&0.486&0.405\\  
18&     &   &   &   &   &   &   &    &   &   &   &   &   &   &   &   &  &1&0.427&0.38&0.483&0.400\\  
19 &   &   &   &   &   &   &   &    &   &   &   &   &   &   &   &   &   &  & 1&0.340&0.427&0.356\\  
20 &    &   &   &   &   &   &   &    &   &   &   &   &   &   &   &   &   &  &   & 1&0.384&0.318 \\  
21  &  &   &   &   &   &   &   &    &   &   &   &   &   &   &   &   &   &  &   &   & 1&0.403\\  
22  & &   &   &   &   &   &   &    &   &   &   &   &   &   &   &   &   &  &   &   &   & 1 \\

\end{tabular}
}
\end{table*}

\section*{Acknowledgments}
We greatly appreciate the efforts of the staff of VEPP-4M to provide
good operation of the complex during long term experiments.
The authors are grateful to V.~P.~Druzhinin for useful discussions.
The Siberian Branch of the Russian Academy of Sciences Siberian
Supercomputer Center and Novosibirsk State University Supercomputer
Center are gratefully acknowledged for providing supercomputer facilities.

This work has been supported by Russian Science Foundation (project N
14-50-00080). SE acknowledges Russian Science Foundation 
(project N 17-12-01036) for supporting part of this work related to 
Monte Carlo generators.


\begin{thebibliography}{99}
\bibitem{dhmz} M.\,Davier {\it et al.},
\newblock Eur.\ Phys.\ J. C {\bf 71}, 1515 (2011).
\bibitem{hlmnt} K.\,Hagiwara {\it et al.},
\newblock J.\ Phys.\ J. G {\bf 38}, 085003 (2011).
\bibitem{quark} N.\,Brambilla {\it et al.},
\newblock Eur.\ Phys.\ J. C {\bf 71}, 1534 (2011).
\bibitem{ADONEMUPI:R1973} M.\,Grilli {\em et~al.},
\newblock   Nuovo\ Cim.\ Lett. A {\bf 13}, 593 (1973).

\bibitem{Mark1:R1977} P.\,A.\,Rapidis  {\it et al.},
 \newblock  Phys.\ Rev.\ Lett. {\bf 39}, 526 (1977).
\bibitem{PLUTO:R1977}   J.\,Burmester {\em et~al.},
\newblock   Phys.\ Lett.\ B  {\bf 66}, 395  (1977).
\bibitem{GG2:R1979}   C.\,Bacci {\em et~al.}, 
\newblock   Phys.\ Lett.\ B {\bf 86}, 234(1979).
\bibitem{MARK2:R1980}  R.\,H.\,Schindler {\it et al.},
\newblock  Phys.\ Rev.\ D {\bf 21}, 2716 (1980).
\bibitem{ADONE:R1981}  B.\,Esposito {\em et~al.},
\newblock  Nuovo\ Cim.\ Lett. {\bf 30}, 65 (1981).

\bibitem{MARK1:R1982}   J.\,L.\,Siegrist  {\em et~al.}, 
\newblock  Phys.\ Lett.\ B {\bf 26}, 969 (1982).
\bibitem{Bai:1999pk}  J.\,Z.\,Bai {\it et al.} (BES Collaboration),
\newblock    Phys.\ Rev.\ Lett. {\bf 84}, 594 (2000).
\bibitem{BES:R2002}   J.\,Z.\,Bai {\em et~al.} (BES Collaboration),
\newblock  Phys.Rev.Lett. {\bf 88}, 101802 (2002).
\bibitem{BES:R2006}  M.\, Ablikim {\em et~al.} (BES Collaboration),
\newblock   Phys.\ Rev.\ Lett.  {\bf 97}, 262001 (2006).
\bibitem{BES:R2009}  M.\, Ablikim {\em et~al.} (BES Collaboration),
\newblock   Phys.\ Lett.\ B {\bf 677},  239 (2009).
\bibitem{KEDR:R2016} V.V. Anashin {\it et.al.} (KEDR Collaboration),  
\newblock   Phys.\ Lett.\ B {\bf 753}, 533 (2016).
\bibitem{KEDR:R2017} V.V. Anashin {\it et.al.} (KEDR Collaboration),  
\newblock   Phys.\ Lett.\ B {\bf 770}, 174 (2017).
\bibitem{Anashin:1998sj} V.\,V.\,Anashin {\em et~al.}, 
\newblock Stockholm 1998, EPAC 98*, 400 (1998).
\bibitem{KEDR:Det}  V.\,V.\,Anashin {\it et.al.} (KEDR Collaboration),
\newblock  Phys. of Part. and Nucl. {\bf 44}, 657 (2013).

\bibitem{psi2S:2012}
  V.\,V.\,Anashin {\it et al.} (KEDR Collaboration),
  Phys.\ Lett.\ B {\bf 711}, 280 (2012).
\bibitem{MASS::KEDR2015} V.\,V.\,Anashin {\em et~al.} (KEDR Collaboration), 
\newblock  Phys.\ Lett.\ B  {\bf 749}, 50 (2015).

\bibitem{PDG:2014}  K.\,A.\,Olive {\em et~al.}~(PDG), 
\newblock  Chin. Phys. C {\bf 38}, 090001 (2014). 
\bibitem{KF:1985} E.\,A.\,Kuraev and V.\,S.\,Fadin,
\newblock   Sov.\ J.\ Nucl.\ Phys. {\bf 41},  466 (1985). 
\bibitem{GEANT:Tool} GEANT -- Detector Description and Simulation
  Tool CERN Program Library Long Writeup W5013.
\bibitem{JETSET} T.\,Sjostrand, M.\,Bengtsson,
\newblock Comp.\ Phys.\ Comm. {\bf 43}, 367 (1987).
\bibitem{LUARLW:2001} Haiming Hu and An Tai,  eConf C010430 (2001) T24,
                      arXiv:hep-ex/0106017.
\bibitem{BHWIDEGEN} S.\, Jadach, W.\, Placzek, B.\,F.\,L.\,Ward,
\newblock \ Phys.\ Lett. B {\bf 390}, 298 (1997).
\bibitem{MCGPJ}  A.\, B.\, Arbuzov {\em et~al.},
\newblock   Eur.\ Phys.\ J. C {\bf 46}, 689 (2006).
\bibitem{KORALB24} S.\,Jadach, Z.\,Was,
\newblock Comp.\ Phys.\ Comm. {\bf 85}, 453 (1995).



\bibitem{BERENDS:EEEE}  F.\,A.\,Berends {\em et al.}, 
\newblock  Comp.\ Phys.\ Comm. {\bf 40}, 285 (1986).
\bibitem{BERENDS:EEMM}  F.\,A.\,Berends {\em et al.}, 
\newblock  Comp.\ Phys.\ Comm. {\bf 40}, 271 (1986).
\bibitem{KEDR:EEX}  V.\,A.\,Tayursky, S.\,I.\,Eidelman, 
\newblock   Preprint IYaF 2000-78, Novosibirsk 2000 (in Russian).
\bibitem{BESGEN} J.\,C.\,Chen {\em et~al.},
\newblock Phys.\ Rev.\ D {\bf 62},  034003 (2000).
\bibitem{jpsi:2018}  V.\,V.\,Anashin {\it et al.} (KEDR  Collaboration),
\newblock   JHEP 1805, 119 (2018), arXiv:1801.01958.
\bibitem{Fox:Wolfram} G.\,C.\,Fox, S.\, Wolfram, 
\newblock Nucl.\ Phys. \ B {\bf 149}, 413 (1979).
\bibitem{Actis:2010gg}
 S.~Actis {\it et al.},
 Eur.\ Phys.\ J.\  C {\bf 66}, 585 (2010).
\bibitem{Fesefeldt:1985yw}   H.\,C.~Fesefeldt,
\newblock  Technical Report PITHA-85-02, III Physilakisches Institut, RWTH Aachen Physikzentrum, 5100 Aachen, Germany, Sep. 1985.
\bibitem{Fasso:2005zz}  A.~Fass$\grave{\text{o}}$  {\em et~al.}, 
\newblock Talk at the Computing in High Energy and Nuclear Physics (CHEP03), arXiv:physics/0306162.
\bibitem{Baikov:pQCD}
  P.\,A.\,Baikov {\em et~al.}, 
\newblock \ Phys.\ Lett. {\bf B 714}, 62 (2012).
\bibitem{Brambilla:2014}
  N.\,Brambilla {\em et~al.}, 
\newblock \ Eur.\,Phys.\, J.\, C {\bf  74},  2981 (2014).
\bibitem{PDG:2018}   M.\, Tanabashi {\em et~al.}~(PDG),
\newblock  \ Phys. \ Rev. {\bf D 98}, 030001 (2018).
\bibitem{mq:2016} J.\,Erler, P.\,Masjuan and H.\, Spiesberger,
\newblock Mod.\, Phys.\, Lett\, A {\bf 31}, 1630041 (2016), arXiv:1611.05648.
\end{thebibliography}
\end{document}